\documentclass[reprint,aps,prl,showpacs,twocolumn,amssymb, amsmath,superscriptaddress,floatfix]{revtex4-2}

\usepackage{docs}%
\usepackage{bm}%
\usepackage{graphicx}
\usepackage[colorlinks=true,linkcolor=blue]{hyperref}%
\begin{document}
\title{Vibration-Assisted Multi-Photon Resonance and Multi-Ion Excitation}
%Multi-Photon Resonance and Multi-Ion Excitation of Hot Trapped Ions
\author{Wenjun Shao}
\affiliation{Department of Physics, Fudan University, Shanghai 200433, China}
\affiliation{Department of Physics, School of Science, Westlake University, Hangzhou 310024,  China}
\affiliation{Institute of Natural Sciences, Westlake Institute for Advanced Study, Hangzhou 310024, China}
% and Institute of Natural Sciences, Westlake Institute for Advanced Study , Westlake University, Hangzhou 310024, Zhejiang Province, P.R. China
\author{Xun-Li Feng}
%\email{xlfeng@shnu.edu.cn}
\affiliation{Department of Physics, Shanghai Normal University, Shanghai 200234,  China}
\author{Jian Li}
\affiliation{Department of Physics, School of Science, Westlake University, Hangzhou 310024,  China}
\affiliation{Institute of Natural Sciences, Westlake Institute for Advanced Study, Hangzhou 310024,  China}
\author{Liang-Liang Wang}
\email{wangliangliang@westlake.edu.cn}
\affiliation{Department of Physics, School of Science, Westlake University, Hangzhou 310024, China}
\affiliation{Institute of Natural Sciences, Westlake Institute for Advanced Study, Hangzhou 310024,  China}

%\date{{\small \today}}  

\begin{abstract}

We investigate the multi-photon resonance and multi-ion excitation in a single-mode cavity with identical vibrating ion-qubits, which enables the tripartite interaction among the internal states of ions, the cavity mode and the ions' vibrational motion.
Under particular resonant conditions, we derive effective Hamiltonians for the three-photon and the three-excitation cases, respectively, and find that the magnitude of the effective coupling energy can be tuned through the vibration mode, allowing for manipulations of ion-photon coupling in experiments. Furthermore, we analyze the system dynamics of our proposed setups and demonstrate the Rabi oscillation behaviors in these systems with dissipation effects. We propose our system as a versatile platform for the exploration of entangled multi-qubit physics.

\end{abstract}

\maketitle

The investigation of resonant emission of multiple photons \cite{Scully,Mayer, Lg} and resonant excitation of multiple atoms \cite{Nori, Nori2, exp1} is important not only due to the fundamental physics involved, but also because of their potential applications in, for examples, the realization of quantum gates and quantum information storage devices.
Following the first realization of two-photon absorption \cite{TP, TP3,TP4,TP5} in the 1960s and its rapid application \cite{2ph1,2ph2,2ph3} as a powerful spectroscopic and diagnostic tool, variant multi-photon processes have attracted growing research interests.
Recently, Ma and Law \cite{Law} found that three photons can be simultaneously absorbed by a two-level atom in the strong coupling (SC) regime of Rydberg atoms confined in a high-Q cavity. On the other hand, Garziano \textit{et al}. \cite{Nori,Nori2} reported a counter-intuitive reverse phenomenon where a single photon can simultaneously excite two or more independent atoms in a symmetry-breaking potential.

Most existing investigations of multi-photon resonance and multi-atom excitation processes assume the atoms to be static with the dipole approximation.
With the development of atom and ion trapping techniques \cite{Wineland0,WPaul,Ashkin}, however, the inclusion of the vibrational degrees of freedom of atoms in a cavity becomes increasingly important.
The trapped ions \cite{ATS,Blatt,Wineland} interacting with a quantized cavity field can be cooled down to their lowest vibrational state \cite{model2}.
In such a cavity quantum electrodynamics (QED) system, the internal ionic states, the quantum vibrational mode of the ion and the single-mode cavity field are necessarily intertwined. For instances, Bu\v{z}ek \textit{et al}. \cite{model2} analyzed the quantum motion of a cold, trapped two-level ion interacting with a quantized light field in a single-mode cavity. At zero temperature, the dynamics of a single trapped ion inside a nonideal QED cavity was studied by Rangel \textit{et al}. \cite{Rangel}.
Nevertheless, the effects of vibrational states in the multi-photon and multi-atom processes remain unexplored to our knowledge.

In this work, we investigate a chain of trapped ions in a single-mode cavity and focus on the quantum effect of ionic vibration on the three-photon resonance and three-ion excitation processes in the cavity-QED system.
We derive effective Hamiltonians close to the resonance points and determine the energy level splittings. This reveals rich physics in our system. Although small vibrational frequency hardly affects the overall energy spectrum, the situation can be completely different and very interesting if the vibrating frequency is comparable to or larger than the ionic transition frequency. Remarkably, the energy splitting can be effectively controlled and enhanced by varying the vibration frequency, particularly around certain resonant conditions, resulting in vibrationally controlled ion-photon coherent manipulation. We further demonstrate the Rabi oscillation dynamics and discuss the associated damping effects, validating our analytical results with concrete numerical simulations.

\begin{figure}[bp]
\centering
\includegraphics[width=0.48\textwidth]{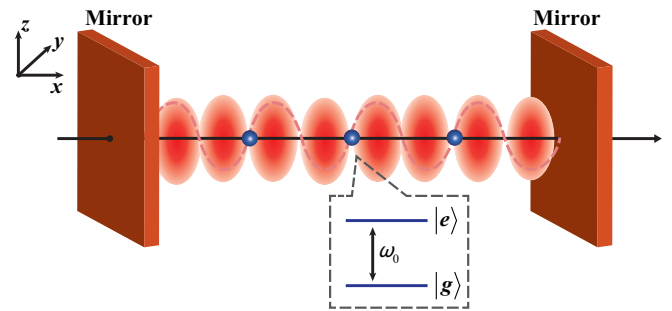} 
\caption{\label{Fig:setup}Schematic diagram of a series of two-level ions trapped inside a standing-wave mode of an optical cavity. All ions are arranged at the nodes of the cavity's standing wave to make sure the tripartite coupling among the atomic internal states, the cavity mode and the vibrational motion to be the only relevant interaction. }
\end{figure}

\textit{Model} --- We consider a chain of $N$ identical ion-qubits  placed in a single-mode high-Q cavity \cite{OC, trappedionC2}. As schematically shown in Fig. \ref{Fig:setup}, the trapped ions are assumed to sit close to the nodes of the cavity field standing wave, which can be best achieved by tuning the longitudinal trapping potential or the wavelength of the cavity field in the few-ion cases ($N\leq 3$). Consequently, the tripartite coupling among the atomic internal states $|e\rangle$ and $|g\rangle$, the cavity mode, and the vibrational motion of the ions, become the only allowed interaction \cite{model2, tripartite, model3, feng}.
We will consider the lowest collective mechanical mode of the ions \cite{collective, CZ, MS}, which is referred to as the center-of-mass mode.
The fully quantized Hamiltonian of the system thus reads (we set $\hbar=1$) \cite{feng, tripartite, model3, OC}
\begin{align}
  &H=H_{0}+H_{\mathrm{int}},\\
  &H_{0}=\nu a^{\dagger}a + \omega_{c}b^{\dagger}b +\frac{1}{2}\omega_{0}J_{z},\\
  &H_{\mathrm{int}}=g\sin[\eta(a^{\dagger}+a)](J_{+}+J_{-})(b^{\dagger}+b),
\end{align}
where $a^{\dagger}$ ($a$) is the creation (annihilation) operator of the ion center-of-mass mode with frequency $\nu$; % of the ion's vibrational motion, 
$b^{\dagger}$ ($b$) is the creation (annihilation) operator of the cavity field mode with the frequency $\omega_{c}$; $J_{+} = \sum_{i}\sigma_{+}^{i} = \sum_{i}|e_{i}\rangle\langle g_{i}|$, $J_{-} = \sum_{i}\sigma_{-}^{i} = \sum_{i}|g_{i}\rangle\langle e_{i}|$,
$J_{z} = \sum_{i}\sigma_{z}^{i} = \sum_{i}(|e_{i}\rangle\langle e_{i}|-|g_{i}\rangle\langle g_{i}|)$ operate on the ionic two-levels with transition frequency $\omega_{0}$; $g$ represents the coupling strength between the cavity mode and the ionic internal states
%, which can be arbitrary as no rotating wave approximation (RWA) is applied, 
and $\eta$ is the Lamb-Dicke parameter. In the case
of strong confinement, the Lamb-Dicke condition $\eta\ll 1$ is naturally satisfied and the interaction Hamiltonian can be transformed conveniently to a trilinear form $H_{\mathrm{int}}\approx g\eta(a^{\dagger}+a)(J_{+}+J_{-})(b^{\dagger}+b)$ with the approximation $\sin[\eta(a^{\dagger}+a)]\approx \eta (a^{\dagger}+a)$.
Note that the high-order resonant transitions that are of our particular interest in this paper can be realized via intermediate states connected by counter-rotating terms (CRTs) in the interaction Hamiltonian, such as $a\sigma_{+}b^{\dagger}$, which describes the creation of a photon in the cavity accompanied by an ionic excitation from its ground state together with the annihilation of a phonon from the ionic vibration.

\begin{figure}[bp]
\centering
\includegraphics[width=0.48\textwidth]{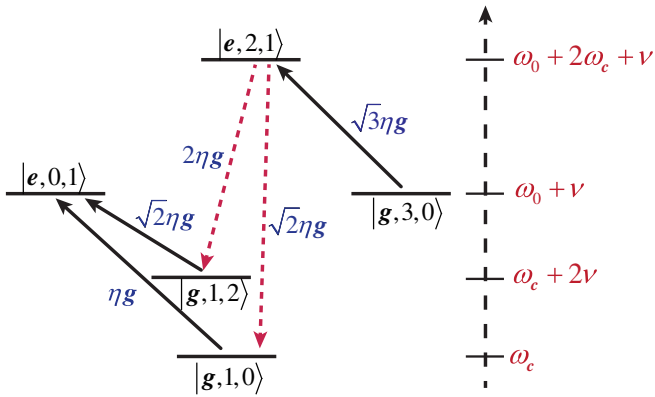} 
\caption{Sketch of two paths contributing to the effective coupling between the bare states $|g,3,0\rangle$ and $|e,0,1\rangle$ via intermediate virtual transitions. The rotating processes are indicated by solid lines and the counter-rotating processes are indicated by dashed lines. The transition matrix elements are also displayed.}
\label{Fig:energy level}
\end{figure}

\begin{figure}[htp]
\centering
\includegraphics[width=0.48\textwidth]{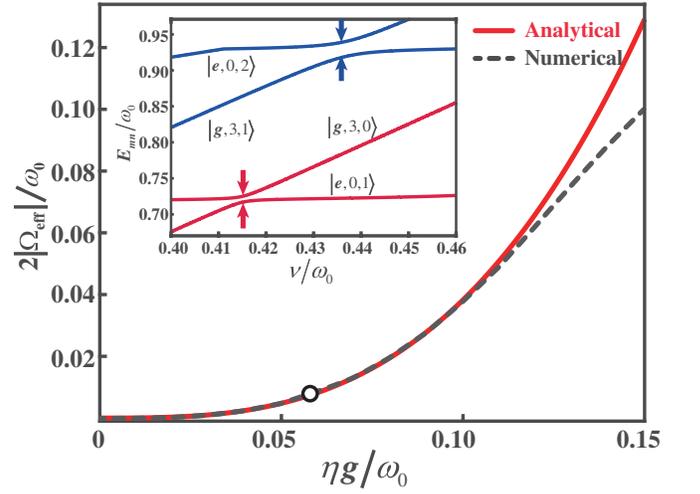} 
\caption{Comparison of the effective energy splitting $2\vert\Omega_{\mathrm{eff}}\vert/\omega_{0}$ obtained analytically (solid-red line) and numerically (dashed line) as a function of interaction strength $\eta g/\omega_{0}$ with $\nu/\omega_{0}=0.2$. The inset shows the energy spectrum for states: $\vert g,3,0\rangle \leftrightarrow \vert e,0,1\rangle$ and $ \vert g,3,1\rangle \leftrightarrow \vert e,0,2\rangle$ at $\eta g/\omega_{0}=0.06$ marked by the black circle \cite{Sup2}. The avoided-level crossings indicated by the arrows occur with the energy splitting about $0.008 \omega_{0}$ and $ 0.021 \omega_{0}$, respectively. The corresponding positions of resonance are $0.415 \omega_{0} $ and $0.437 \omega_{0} $.}
\label{Fig:Energy Splitting}
\end{figure}

%Variable/tunable
\textit{Energy splitting controlled by vibrational mode} --- For the sake of simplicity, we first investigate the single ion case with $N=1$. Under the resonant condition with $\omega_{c}\approx (\omega_{0}+\nu)/3$, the dominant coupling terms are $a^{\dagger}b^3\sigma_{+}$ and $a(b^{\dagger})^{3}\sigma_{-}$, describing the ion excited from its ground state by annihilating three photons while creating one phonon, as well as its inverse process. 
As illustrated in Fig.  \ref{Fig:energy level}, the presence of CRTs in the interaction Hamiltonian enables two different paths for the transitions $\vert g,3,0\rangle \leftrightarrow \vert e,0,1\rangle$. By applying standard third-order perturbation theory \cite{QM,exp1}, we obtain the effective coupling rate between bare states $\vert g,3,0\rangle$ and $\vert e,0,1\rangle$, under the resonant condition $3\omega_c =  \omega_{0}+\nu $, to be 
\begin{equation}
\Omega_{\mathrm{eff}}= \frac{27\sqrt{6}(\eta g)^{3}\omega_{0}}{4(\omega_{0}+\nu)^2(\omega_{0}-2\nu)}.
\label{Eq:effective coupling with N=1}
\end{equation}
The effective interaction Hamiltonian of interest is
$H_{\mathrm{eff}}=-\Omega_{\mathrm{eff}}\left(\vert e,0,1\rangle \langle g,3,0 \vert +H.c. \right)$. 
Typically, $2\Omega_{\mathrm{eff}}$ can be understood as the energy splitting at the avoided crossing (see Fig.~\ref{Fig:Energy Splitting}), which originates from the hybridization of the states $|g,3,0\rangle$ and $|e,0,1\rangle$.

\begin{figure}[htp]
%\centering
%\includegraphics[width=0.5\textwidth]{3DOmega.png} 
\includegraphics[width=0.5\textwidth]{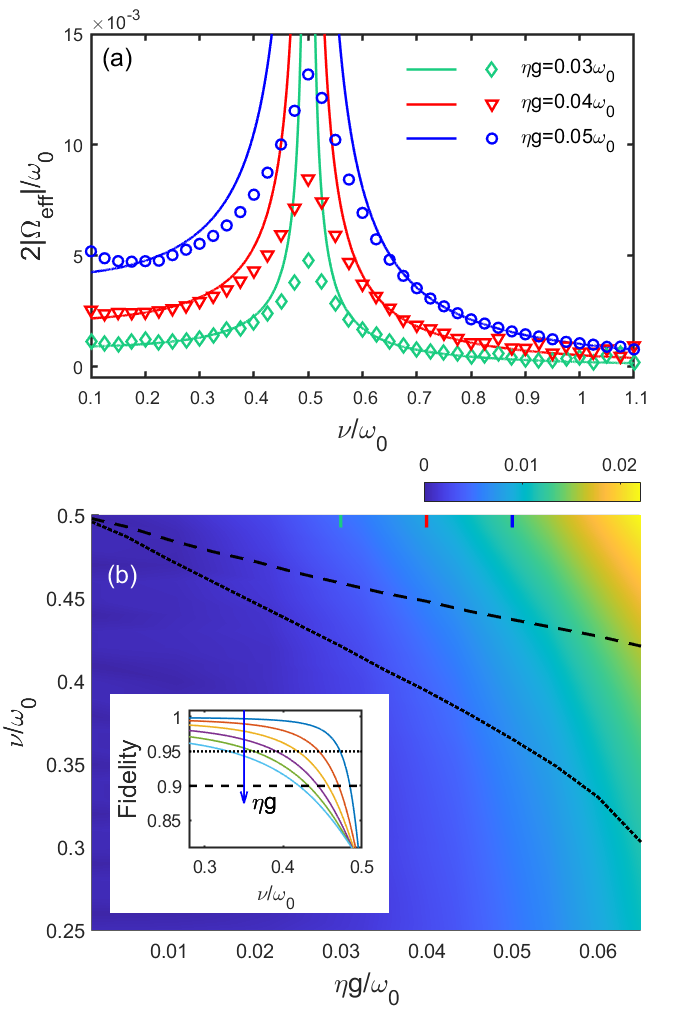} 
\caption{(a) Energy splitting $2\vert \Omega_{\mathrm{eff}}|/\omega_{0}$ of $\vert g,3,0\rangle \leftrightarrow \vert e,0,1\rangle$ obtained numerically as a function of $\nu/\omega$ for $\eta g =0.03\omega_{0}$ (green diamond), $\eta g =0.04\omega_{0}$ (red triangle), and $\eta g =0.05\omega_{0}$ (blue circle), as well as the corresponding analytical results (solid lines). Note that the numerical results show finite peaks at the analytically divergent point at $\nu/\omega_{0}=0.5$.
(b) The numerically-calculated resonant energy splitting in the $\nu-\eta g$ plane. 
The inset shows the %systematic 
fidelity, defined as the average probability weights on the states $\left\vert g,3,0\right\rangle $ and $\left\vert e,0,1\right\rangle $, under the resonant condition for  $\eta g$ varying from $0.01\omega_{0}$ to $0.06 \omega_{0}$ (lines from top to bottom). The 90\% and 95\%  fidelity levels are indicated in both the main plot and the inset by the dashed and the dotted lines, respectively. In the main plot, the three values of $\eta g$ considered in panel (a) are also marked with short line segments of corresponding colors.}
\label{Fig:3DOmega}
\end{figure}
  
% The fidelity equals to 90\% and 95\% marked in dashed line and dotted line, respectively. 

Clearly, when $\nu$ is small compared with the ionic transition frequency $\omega_0$, %the Rabi splitting ratio 
$2\Omega_{\mathrm{eff}}/\omega_{0}$ is proportional to the cubic of the coupling coefficient, i.e., $2\Omega_{\mathrm{eff}}/\omega_{0}\sim (\eta g)^3/\omega_{0}^3$. Therefore, in order to observe this energy splitting, the ultra-strong coupling (USC)  \cite{ultra,ultra2} or even deep-strong coupling (DSC) regime \cite{DSC} is required to ensure that the effective coupling induced by the higher-order processes becomes larger than the relevant decoherence rate in the system.
In Fig. \ref{Fig:Energy Splitting}, we further compare analytically obtained Eq.~\eqref{Eq:effective coupling with N=1} with the energy splitting $2\Omega_{\mathrm{eff}}/\omega_{0}$ obtained from numerically diagonalizing the original Hamiltonian at different $\eta g/\omega_{0}$. % to compare the analytical result demonstrated by the solid curve. 
The results from the two methods agree very well at relatively small coupling (the percentage difference is lower than $2\%$ for $\eta g/\omega_{0}< 0.1$), validating our perturbation theory formula Eq.~\eqref{Eq:effective coupling with N=1} as a good approximation. %(the percentage difference is lower than $0.5\%$ for $\eta g/\omega_{0}< 0.06$)
We note that the above results can be generalized to other pairs of bare states $|g,m,n\rangle$ and $|e,m-3,n+1\rangle$, where $m$ $ (n) $ denotes the photon (phonon) number, and the energy splitting can be formally enhanced with a larger number of photons or phonons (see Supplemental Material  \cite{Sup2}). %The cavity field at which the resonance occurs is determined by simply equaling diagonal elements of $\hat{H}_{\mathrm{eff}}$, and a solution of the resonance frequency  (shown in Sec. I of \cite{Sup2})

When the vibrational frequency is comparable to or larger than the ionic transition frequency and the cavity mode frequency, the picture can be complicated by  the additional longitudinal modes of the cavity generated due to the rapid ionic vibration \cite{feng}. In this regard, the capability of the cavity to maintain a single mode \cite{OC} is particularly useful for us to neglect such a complication in our following discussion.
To achieve a more prominent energy splitting, we consider the vibrational frequency in the regime $0< \nu < \omega_{0}$. As shown in Fig. \ref{Fig:3DOmega}(a), the perturbative result in Eq. \eqref{Eq:effective coupling with N=1} suggests a divergence of the effective coupling $\Omega_{\mathrm{eff}}$ at $\nu/\omega_{0}= 1/2$, which is regularized in the numerical calculation as a pronounced but finite peak. This divergence point is indeed where the perturbation theory breaks down because of the occurrence of higher level degeneracy involving other states than the two original states in our problem. Specifically, for the case with the initial state $\left\vert g,3,0\right\rangle$ or $\left\vert e,0,1\right\rangle$, the presumed intermediate state $\left\vert g,1,2\right\rangle $ becomes energetically aligned with these two states in the decoupled limit when $\nu/\omega_{0}= 1/2$, such that the coupling will lead to hybridization of all three states and both our perturbation theory and the two-level resonance picture will fail close to this point. Hence we introduce a fidelity measure for the two-state transition, which is defined by the total probability weights of the two relevant states $\left\vert g,3,0\right\rangle $ and $\left\vert e,0,1\right\rangle$.
In Fig. \ref{Fig:3DOmega}(b), we show the line contours for where this fidelity is equal to $90\%$ (the dashed line) and $95\%$ (the dotted line), respectively. By following such contours, optimized values of the interaction strength $\eta g$ and the vibrational frequency $\nu$ can be found with a specific fidelity threshold.
It is quite interesting that the energy splitting can be greatly enhanced by the ionic vibration mode, making high-order resonant transitions significant for realistic experimental parameters. As we will see in below, such enhanced energy splitting also appears in the system with multiple trapped ions.

\textit{Vibration-assisted multi-ion excitation process} --- We now turn to study a chain of $N=3$ ions and discuss a specific multi-ion excitation process: one photon and one phonon can be jointly absorbed by three ions.
Following similar procedures as in the above section and utilizing the generalized James' effective Hamiltonian method \cite{james}, we arrive at 
\begin{equation}
H_{\mathrm{eff}}=-\Omega_{\mathrm{eff}}[ab\sigma_{+}^{1}\sigma_{+}^{2}\sigma_{+}^{3}+a^{\dagger}b^{\dagger}\sigma_{-}^{1}\sigma_{-}^{2}\sigma_{-}^{3}],
\label{EG:Heff3}
\end{equation} 
with 
\begin{equation}
\frac{\Omega_{\mathrm{eff}}}{\omega_{0}}=\frac{3\nu(3\omega_{0}-\nu)}{2(\omega_{0}-\nu)(2\omega_{0}-\nu)} \left( \dfrac{\eta g}{\omega_{0}}   \right)^3 
\label{Eq:effective coupling with N=3}
\end{equation}
being the effective coupling strength  for the transition between the states $|ggg,1,1\rangle$ and $|eee,0,0\rangle$ \cite{Sup2}, where the resonance condition $\omega_{c}=3\omega_{0}-\nu$  has been taken. %with $\nu/\omega_{0}\ll1$
The above result resembles that in the case of simultaneous excitation of three ions by only one photon, except that in the latter case the strength of the effective coupling vanishes on resonance $\omega_{c}=3\omega_{0}$ (discussed in Ref. \cite{Nori} and the case IV B 2c of Ref. \cite{exp1}) because of the destructive interfere between different transition paths. By contrast, the energy splitting will be finite on resonance as long as the transition is assisted by a vibration mode (see Sec. III in Supplemental Material \cite{Sup2}). In addition, similar to the preceding three-photon resonance process, the energy splitting implied by Eq. \eqref{Eq:effective coupling with N=3} shows two divergences at $\nu/\omega_{0}=1,2 $, which indicate the breaking down of the perturbation theory and are regulated in the numerical results as two pronounced peaks (see Fig. \ref{Cp113} in Supplemental Material \cite{Sup2}).

\textit{System dynamics} --- In order to better demonstrate the processes we have proposed, we now describe the system dynamics with all the dissipation channels taken into account.
Here, we adopt the master equation approach following Ref. \cite{Lg, Nori, Nori2, Dissipation, Yin} as the standard quantum optical master equation breaks down in the USC regime. With the Born-Markov approximation and assuming the system interacting with zero-temperature baths, the Lindblad master equation regarding our system is given by
\begin{equation}
\begin{split}
\dot{\rho}(t)= &-i\left[H,\rho(t)\right]+\kappa \mathcal{D}[X^+]\rho(t) \\
&+\gamma \sum_i \mathcal{D}[C_i^+]\rho(t)+\zeta \mathcal{D}[P^+]\rho(t),
\end{split}
\end{equation}
where the Lindblad superoperator $ \mathcal{D} $ is defined as
$ \mathcal{D}[O]\rho=\dfrac{1}{2}(2O\rho O^{\dagger}-\rho O^{\dagger}O-O^{\dagger}O\rho) $ with $ O=\sum_{j,k>j} \langle j \vert (o+o^{\dagger})\vert k\rangle  \vert j\rangle \langle k \vert$ being the dressed lowering operator (positive frequency part) for the cavity field ($o=b$, $O  = X^+$), the vibrational mode ($o=a$, $ O  = P^+$) and the $ i$-th ion ($o=\sigma_-^i$, $ O  = C_i^+$), respectively. 
Correspondingly for the negative frequency part, $ O^{\dagger} = X^-$,  $P^-$ or $ C_i^-$. The constants $ \kappa $,  $\zeta$  and $ \gamma $ correspond to the damping rates for the cavity mode, the vibration mode and the ions, respectively.

\begin{figure}[htp]
\centering
\includegraphics[width=0.495\textwidth]{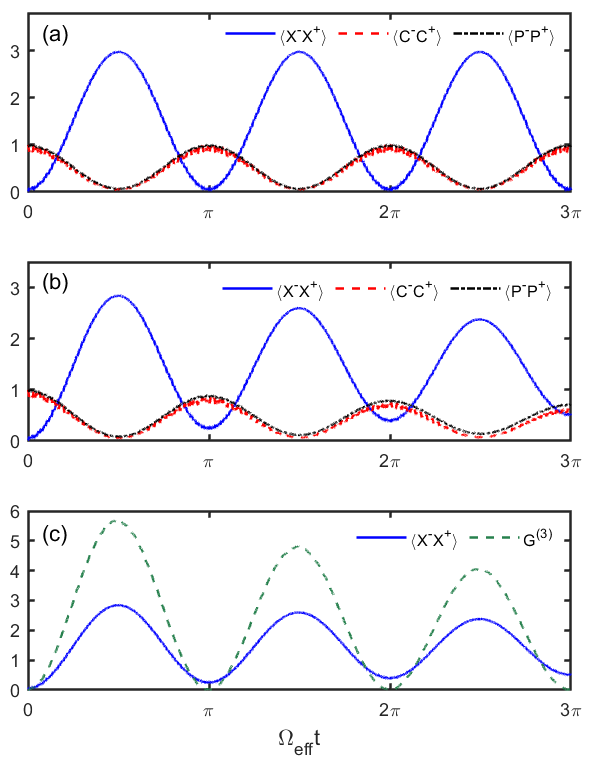}
\caption{Time evolution of the ion mean excitation number $ \langle C^- C^+ \rangle $ (red dashed curve), the cavity mean photon number $ \langle X^- X^+ \rangle $ (blue solid curve), the three-photon correlation function $ G^{(3)}$ (green dashed curve) and the mean phonon number $ \langle P^- P^+ \rangle $ (black dot-dashed curve) with (a) no decay and (b,c) decay rates $ \kappa=\gamma= 2\zeta=1\times10^{-4}\omega_0 $, respectively. The initial state is taken to be $ \left\vert e,0,1\right\rangle $ and the other parameters are  $\eta g/\omega_{0}=0.06$ and $\nu/\omega_{0}=0.15$.}
\label{TE311C}
\end{figure}

For the case of three-photon resonance, the system is initially prepared in the state $ \left\vert e,0,1\right\rangle $.  Fig. \ref{TE311C} shows the time evolution of the ion mean excitation number $ \langle C^- C^+ \rangle $, the cavity mean photon number $ \langle X^- X^+ \rangle $, the three-photon correlation function $ G^{(3)}= \langle X^- X^- X^- X^+ X^+X^+ \rangle$ and the mean phonon number $ \langle P^- P^+ \rangle $. In the ideal case without dissipation (Fig. \ref{TE311C}(a)),  the mean photon number at its maximum approaches $3$ and meanwhile the ion transits to its ground state and the mean phonon number approaches $ 0 $, which is an signature of the cavity mode being in a three-photon state excited by the ion and one phonon. 
This process is reversible accompanied by energy exchanges. 
When the dissipation effect is included in the system dynamics [Fig. \ref{TE311C}(b)], the mean values oscillate with the time evolution but decrease exponentially in their amplitudes as expected. We observe in Fig. \ref{TE311C}(c) that the peak values of $ G^{(3)}$ are approximately twice of the mean photon number $ \langle X^- X^+ \rangle $ in the first transition cycle, indicating an almost-perfect three-photon correlation \cite{Lg}.
Similarly, the Rabi oscillation of multi-ion excitation also shows a reversible energy exchange with the decay rate of its amplitudes depending on the relative strength between the effective coupling and the system loss (see details in Supplemental Material  \cite{Sup2}).

\textit{Discussion on experimental implementations} ---  For the case we have discussed with $N=1,3$, the ions are assumed to be distributed uniformly \cite{trappedionC, equal,equal3} in a Radio-Frequency (RF) linear Paul trap \cite{trap1}. The ions are strongly bound in radial %the $y$ and $z$ 
directions and weakly bound in an harmonic potential in the axial direction.
The equilibrium position of the ion is determined by the joint effect of the overall harmonic trap and the Coulomb force, 
and the ion-ion distance can be tuned via the trap frequency.
To achieve the vibration frequency of the ion comparable to the cavity field frequency and ionic transition frequency, a microwave cavity with resonant frequency in the $\mathrm{MHz}$ spectral range is adopted and a suited Rydberg transition of $\mathrm{Rb}$ ion can be achieved \cite{21h}, which is feasible with current existing techniques.
A noteworthy issue is that with increasing number of ions ($N>3$), the ion distribution in a RF trap becomes inhomogeneous and the ion-ion distance increases from the trap centre to the edges. For a system involving only a small number of ions such as in our proposals, however, we expect the ion-ion distance only varies insignificantly and the equilibrium positions of the ions are still close enough to the nodes of the cavity standing wave. Recently, a novel technique \cite{equal3,homogeneous} has been developed to realize an uniform ion distribution in a long range by controlling the electrode voltage of a surface ion trap, which may also be adapted for implementations of our proposals.

\textit{Conclusion} --- We have investigated the cavity QED with vibrating two-level ions in the large-detuning regime, focusing on two phonon-assisted phenomena: three-photon resonance and three-ion simultaneous excitation.
Such cavity QED systems can be used to prepare complex entangled states such as the N00N states and the Greenberger-Horne-Zeilinger (GHZ) states (see Supplemental Material  \cite{Sup2}), or to realize three-body effective interaction \cite{3body, 3body2, 3body3} among trapped ions.
With the state-of-the-art techniques in optical lattices, circuit QED systems and optomechanical systems \cite{cavityom,cavityom2}, we expect our proposals can be generalized to \cite{Sup2} various experimental platforms with coupling between qubits and multiple types of bosons such as resonator modes, optical or acoustic phonons \cite{SAW}, thus providing more insights into quantum entanglement physics and promoting the further development of novel quantum technology.  \newline

We would like to thank Chunfeng Wu, Jinxing Hou, Yong Sun and Chang-An Li for interesting discussions. This work was supported by foundation of Zhejiang Province Natural Science under Grant No. LQ20A040002. JL acknowledges support from National Natural Science Foundation of China under Project 11774317. 
%The numerical calculations in this paper have been done on the super-computing system in the Information Technology Center of Westlake University.

\clearpage

\begin{widetext}
\subsection*{ {\Large Supplemental Material for ``Vibration-Assisted Multi-Photon Resonance and Multi-Ion Excitation"} }

\subsection*{I. Model generalization }
% mentioned in Ref.
With the rapid development in circuit QED systems, optomechanical systems and optical lattices, our model can be further generalized to various experimental platforms by coupling qubits to two bosonic modes, such as resonator modes, optical or acoustic phonons. 
For example, the ion vibrational mode can be replaced by the phonon mode of a mechanical resonator in Hybrid Optomechanical Systems \cite{cavityom02}. The interaction between two-level atoms and an optical cavity can be described by Rabi model through the Hamiltonian
\begin{equation}
H_{\mathrm{int}}=g (J_{+}+J_{-})(b^{\dagger}+b),\label{Rabi} \tag{S1}
\end{equation}
where $ g= - \textbf{d} \cdot \varepsilon_0 $ is the atom-photon coupling rate. 
Consider a standard dispersively coupled optomechanical system with the frequency of a mechanical resonator $ \nu $  and phonon annihilation operator $ a $, 
in which the resonator's displacement $ x=x_{zpf}(a+a^{\dagger}) $ ($ x_{zpf} $ is the zero-point motion amplitude) creates  spatial influence on the cavity field (see more details in Ref. \cite{cavityom02}). 
Thus, the atom-photon coupling rate $ g $ becomes dependent on the mechanical position.  
Expanded to first order in $ x $, $ g(x)=g(0)+\gamma(a+a^{\dagger}) $, where $\gamma=(\partial g/\partial x) \vert_{x=0} x_{zpf} $. 
Inserting $ g(x)$ into the Hamiltonian \eqref{Rabi} and taking $ g(0)= 0$, %that is, at mechanical equilibrium the field at the emitter's position vanishes,
a novel atom-photon-phonon coupling arise, which is termed \textit{mode field coupling} (MFC).
The interaction Hamiltonian is
\begin{equation}
H_{\mathrm{int}}=\gamma  (a+a^{\dagger}) (J_{+}+J_{-})(b^{\dagger}+b),\label{Hmfc} \tag{S2}
\end{equation}
which is the only possible interaction allowing the swap of excitation between three quantum systems.  This effective MFC interaction Hamiltonian [Eq. \eqref{Hmfc}] from hybrid optomechanical systems can also realise three-photon resonance and three-ion excitation.

\newpage

\subsection*{II. Transition paths and the effective Hamiltonian} 
\subsubsection{A. Phonon-assisted three-photon resonance}

According to standard perturbation theory \cite{Nori02, QM2, petb2}, the magnitude of the effective coupling strength is  
\begin{equation} 
\Omega =\sum_{j_{1},j_{2}\cdots j_{n-1}}\frac{V_{fj_{n-1}}\cdots V_{j_{2}j_{1}}V_{j_{1}i}}{(E_{i}-E_{j_{1}})(E_{i}-E_{j_{2}})\cdots(E_{i}-E_{j_{n-1}})},\label{Petb} \tag{S3}
\end{equation} 
where $E_{j_{n}}$ represents the energy of the bare state $ \left\vert j_{n}\right\rangle $, while $ V_{j_{n}j_{n+1}}=\left\langle j_{n}\right\vert  H_{int}\left\vert j_{n+1}\right\rangle $. The sum goes over all of the virtual transition steps which forms a transition path connecting the initial state $ \left\vert i\right\rangle $ to the final state $ \left\vert f\right\rangle $. 

Specifically, we investigate such a case \textbf{(\emph{a})}: one ion is excited from its ground state by annihilating three photons and one phonon. The state $ \left\vert g,3,1\right\rangle  $, besides its self-energy, can connect itself via second-order perturbation, as well as for state $ \left\vert e,0,0\right\rangle$, which have been shown in Fig. \ref{p311a}(a). 
Between the initial state $ \left\vert g,3,1\right\rangle $ and the final state $ \left\vert e,0,0\right\rangle$, we find two  paths and each path includes three virtual transitions, which is shown in Fig. \ref{p311a}(b). 
%Higher-order processes can also contribute, but with little contribution. %occur with lower possibility. 
\begin{figure}[bp]
\renewcommand\thefigure{S1}
\centering
\includegraphics[scale=0.85]{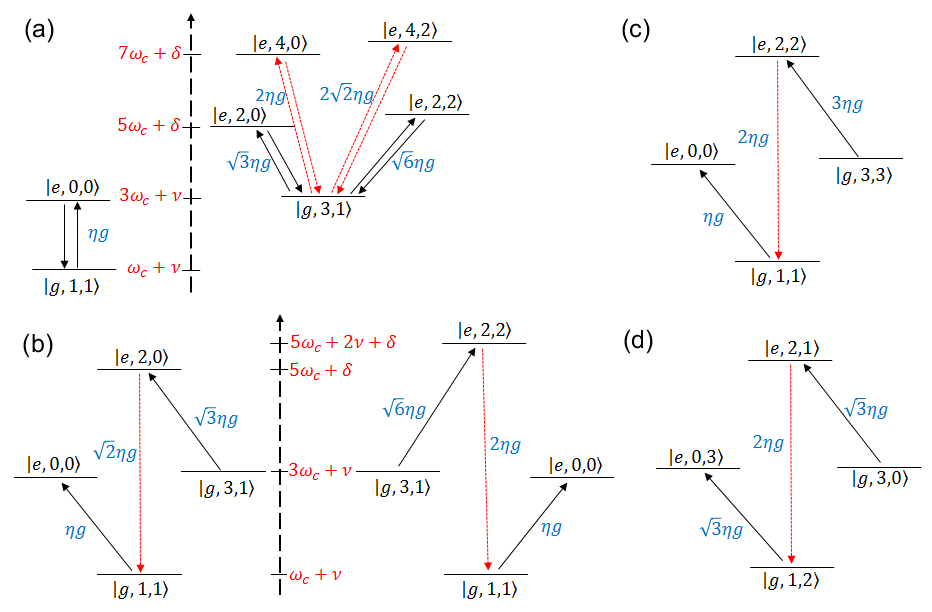}
\caption{(a) Sketches of the processes giving the main contribution to energy shift for the bare states $ \left\vert g,3,1\right\rangle  $ and $ \left\vert e,0,0\right\rangle$. Coupling between the bare states (b) $ \left\vert g,3,1\right\rangle  $ and $ \left\vert e,0,0\right\rangle$, (c) $ \left\vert g,3,3\right\rangle  $ and $ \left\vert e,0,0\right\rangle$, (d) $ \left\vert g,3,0\right\rangle  $ and $ \left\vert e,0,3\right\rangle$ via intermediate virtual transitions. }
\label{p311a}
\end{figure}
By applying  third-order perturbation theory with $ \omega_{c}=(\omega_{0}-\delta)/3$, where the detuning $ \delta $ is very close to $ \nu $,  
we obtain the following effective  Hamiltonian
\begin{equation}\label{Heffa} \tag{S4}
\begin{split}
H_{\mathrm{eff}}=& \left[3\omega_c+\nu-\dfrac{\omega_0}{2}-3(\eta g)^{2} \left( \dfrac{3}{\omega_0+2\nu} + \dfrac{4}{2\omega_0+\nu}+ \dfrac{5}{2(\omega_0-\nu)} \right)  \right] \left\vert g,3,1\right\rangle  \left\langle g,3,1\right\vert \\
&+\left[\dfrac{\omega_0}{2}+\dfrac{3(\eta g)^{2}}{2(\omega_0-\nu)} \right] \left\vert e,0,0\right\rangle  \left\langle e,0,0\right\vert - \Omega^{(3a)}_{\mathrm{eff}} \left(  \left\vert g,3,1\right\rangle  \left\langle e,0,0\right\vert + \left\vert e,0,0\right\rangle  \left\langle g,3,1\right\vert\right), 
\end{split}
\end{equation}
with $\Omega^{(3a)}_{\mathrm{eff}} =\frac{27\sqrt{6}\omega_{0}(\eta g)^{3}}{4(\omega_{0}-\nu)^{2}(\omega_{0}+2\nu)}$ \cite{deng2}. 
In Fig \ref{Ens311}(a), we plot the energy spectrum as a function of $ \omega_c $ when $ \nu/\omega_0 =0.2$. In the region around $ \omega_c/\omega_0\approx 0.3 $, there is an avoided crossing between $ \left\vert g,3,1\right\rangle  $ and $ \left\vert e,0,0\right\rangle$, which serves as a signature of a resonance \cite{Law02}. 
In the effective Hamiltonian \eqref{Heffa}, $ 2 \Omega^{(3a)}_{\mathrm{eff}} $ indicates the energy splitting at the avoided crossing.
We compare this analytical result with the energy splitting to check the validity of $ \Omega^{(3a)}_{\mathrm{eff}}$. The results are shown in Fig. \ref{Cpr}(a) as a function of $\eta g /\omega_0$ with $\nu /\omega_0= 0.2$. It can be seen that $ \Omega^{(3a)}_{\mathrm{eff}}$  agrees well with the numerical results. For instance, the percentage difference is less than $5\% $ for $\eta g /\omega_0 <0.03$. 
Fig. \ref{Cpr}(b) displays the comparison of the absolute value of $ 2 \Omega^{(3a)}_{\mathrm{eff}} $ obtained analytically and numerically as a function of  $\nu/\omega_{0} $ with $ \eta g/\omega_0 =0.06$. 
When $ \nu/\omega_0 =1$, the value of $2 \vert \Omega^{(3a)}_{\mathrm{eff}} \vert$ shows a divergence, which is conformed by a pronounced peak of the numerically-calculated energy splitting occurring at the same $\nu/\omega_{0} $. 
As $ \nu /\omega_0 $ gets close to the divergence point, the degeneracy of multiple states occur, where  perturbation theory breaks down and no longer applies. 
To get a more exact and reliable value, it is useful to take a specific fidelity threshold. 

\begin{figure}[btp]
\renewcommand\thefigure{S2}
\centering
\includegraphics[width=7 in]{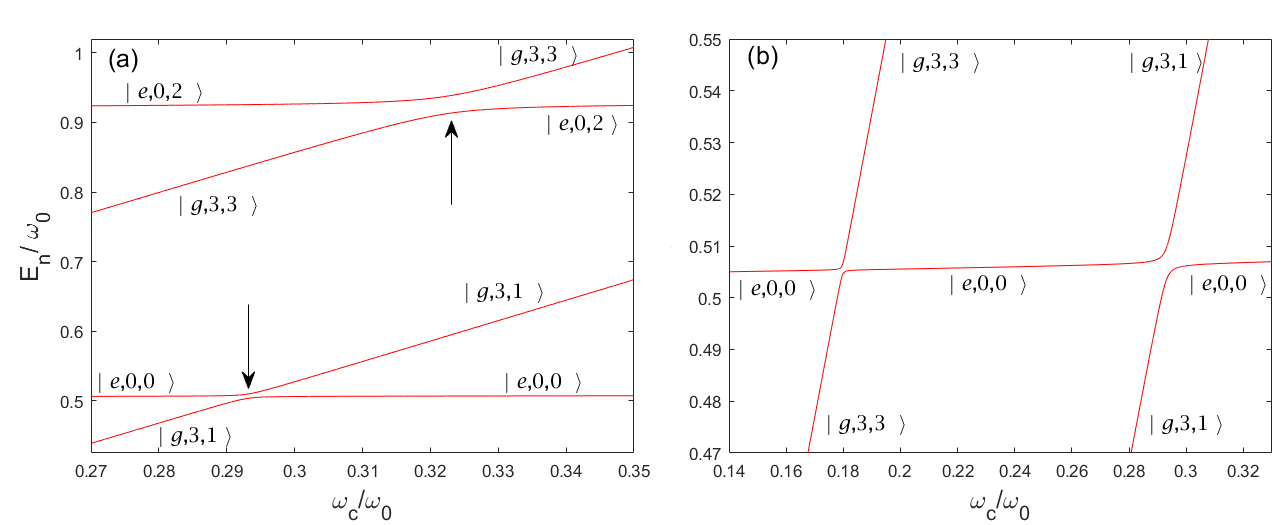}
\caption{Plots of $ E_n/\omega_0 $  as a function of $ \omega_c/\omega_0 $ when $\nu /\omega_0= 0.2$, $\eta g /\omega_0= 0.06$. (a) The resonant field frequency for the states $ \left\vert g,3,1\right\rangle  $ and $ \left\vert e,0,0\right\rangle $ is given by $ \omega_c' \approx 0.293 \omega_0$, at which the avoided crossing of the energy levels occurs with the magnitude of the energy splitting about $ 6.19\times 10^{-3} \omega_0$, and the energy splitting is $ 0.025 \omega_0$ for the states $ \left\vert g,3,3\right\rangle  $ and $ \left\vert e,0,2\right\rangle $ at  $ \omega_c'' \approx 0.322 \omega_0$. (b) At $ \omega_c \approx 0.18 \omega_0$, the avoided-crossing demonstrates the coupling between the states $ \left\vert g,3,3\right\rangle $ and $ \left\vert e,0,0\right\rangle $ with the magnitude of the energy splitting about $1.987\times10^{-3} \omega_0 $.}
\label{Ens311}
\end{figure}

\begin{figure}[btp]
\renewcommand\thefigure{S3}
\centering
\includegraphics[width=5.5 in]{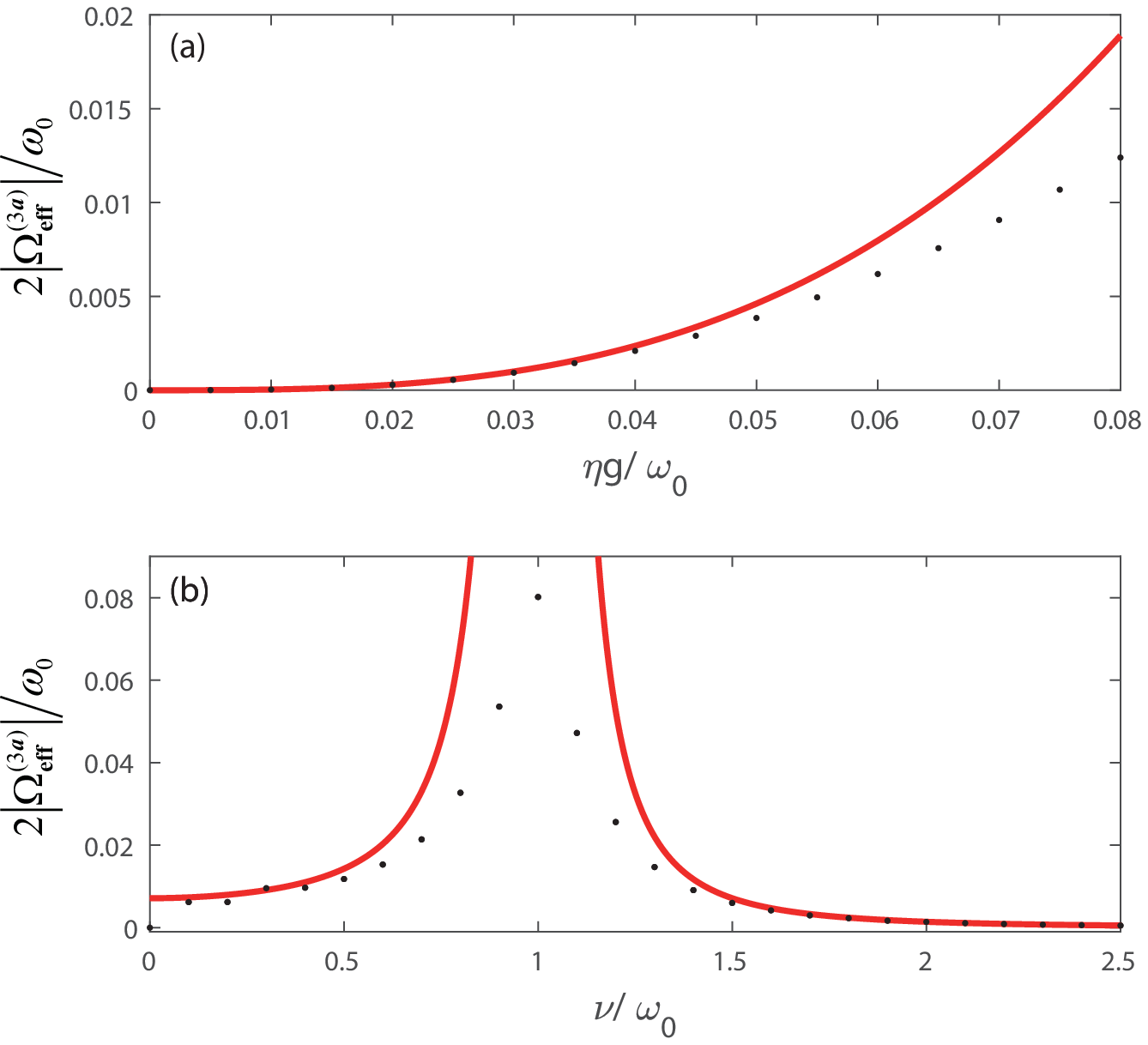}
\caption{(a) Comparison of the magnitudes of the energy splitting $ 2\Omega_{\mathrm{eff}}/\omega_0 $ obtained analytically (solid red line) and numerically (black points) as a function of the interaction strength $ \eta g /\omega_0 $ with $ \nu/\omega_0 =0.2$. (b) Plot of $2\vert \Omega_{\mathrm{eff}} \vert/\omega_0$ in red line as a function of $ \nu/\omega_0 $  and the energy splitting obtained numerically in black points when $ \eta g/\omega_0 =0.06$.
}
\label{Cpr}
\end{figure}
Moreover,  same results can be achieved by applying the generalized James' effective Hamiltonian method \cite{james02}, the general form of the effective Hamiltonian is given by \cite{Heffa2}		
\begin{align}
H_{\mathrm{eff}}&=H_0+ H^{(2a)}_{\mathrm{eff}}+H^{(3a)}_{\mathrm{eff}},\label{Ha}\tag{S5}\\
H^{(2a)}_{\mathrm{eff}}&=V^{(2a_1)}_{\mathrm{eff}} \sigma_+\sigma_- -V^{(2a_2)}_{\mathrm{eff}}\sigma_-\sigma_+,\label{H2a}\tag{S5a}\\
H^{(3a)}_{\mathrm{eff}}&=-V^{(3a)}_{\mathrm{eff}} \left[ anb^3\sigma_+ + na^{\dagger}(b^{\dagger})^3\sigma_- \right],\label{H3a}\tag{S5b}
\end{align}
where $V^{(2a_1)}_{\mathrm{eff}} =\dfrac{3(\eta g)^{2}}{2} \left[\dfrac{n(m+1)}{\omega_0+2\nu} +\dfrac{nm}{2\omega_0+\nu}+\dfrac{(n+1)(3m+2)}{2(\omega_0-\nu)} \right] $, $V^{(2a_2)}_{\mathrm{eff}}=\dfrac{3(\eta g)^{2}}{2} \left[\dfrac{(n+1)m}{\omega_0+2\nu}+  \dfrac{(n+1)(m+1)}{2\omega_0+\nu} \right.$  $ \left. +\dfrac{n(3m+1)}{2(\omega_0-\nu)}\right] $ and  $V^{(3a)}_{\mathrm{eff}}=\frac{27\omega_{0}\left(\eta g\right)^{3}}{4\left(\omega_{0}-\nu\right)^{2}\left(\omega_{0}+2\nu\right)}  $ with $ n=a^{\dagger}a $, $ m=b^{\dagger}b $. The effective coupling between $ \left\vert g,3,1\right\rangle  $ and $ \left\vert e,0,0\right\rangle$ is $\Omega^{(3a)}_{\mathrm{eff}}=\sqrt{6} V^{(3a)}_{\mathrm{eff}}$. 
The resonant frequency  can be determined from the effective Hamiltonian \eqref{Heffa} or \eqref{Ha}. By equating the diagonal elements of $H_{\mathrm{eff}}$ \cite{Diag}, a solution of the cavity field frequency 
\begin{equation}\label{Res} \tag{S6}
\dfrac{\omega'_c}{\omega_0}=\dfrac{1}{3}\left(1-\dfrac{\nu}{\omega_0}\right)+\left(\dfrac{4}{2+\dfrac{\nu}{\omega_0}}+\dfrac{3}{1+2\dfrac{\nu}{\omega_0}}+\dfrac{3}{1-\dfrac{\nu}{\omega_0}} \right)\left(\dfrac{\eta g}{\omega_0}\right)^2 + O\left(\dfrac{\eta g}{\omega_0}\right)^4 
\end{equation}
is achieved. The $ \left(\eta g\right)^2  $  dependence is due to the Stark shifts of energy levels \citep{Law02}. We have also compared the resonance frequency predicted by Eq. \eqref{Res} with the numerical results. For  $ \eta g /\omega_{0}<0.04 $, the percentage difference is less than 0.1\% when $ \nu /\omega_{0}=0.2 $. 
It is worth noting that the three-photon-one-phonon coupling scheme can be generalized to other pairs of bare states in the form $ \left\vert g,m+3,n+1\right\rangle  $ and $ \left\vert e,m,n\right\rangle $ with energy splitting being $ 2(n+1)\sqrt{(n+1)(m+1)(m+2)(m+3)}\Omega^{(3a)}_{\mathrm{eff}} $. Clearly, with the increase of the photon or phonon number, the magnitude of the energy splitting is increased \cite{Law02},  which is shown in Fig. \ref{Ens311}(a). 
Correspondingly,  the position of the resonance presented in Fig. \ref{Omegc} is
\begin{equation}\label{Resnm} \tag{S7}
\dfrac{\omega'_c(n,m)}{\omega_0}=\dfrac{1}{3}\left(1-\dfrac{\nu}{\omega_0}\right)+\left[\dfrac{5\dfrac{\nu}{\omega_0}+4}{\left(2+\dfrac{\nu}{\omega_0}\right)\left( 1+2\dfrac{\nu}{\omega_0}\right)}  + \dfrac{3(n+1)(m+2)\left(5\dfrac{\nu}{\omega_0}+4\right)}{2\left(1-\dfrac{\nu}{\omega_0}\right)\left(2+\dfrac{\nu}{\omega_0}\right) \left(1+2\dfrac{\nu}{\omega_0}\right)} \right]\left(\dfrac{\eta g}{\omega_0}\right)^2, 
\end{equation}
which also shifts with photon and phonon number.

Eq. \eqref{Ha} just describes the situation when $ 0<\nu / \omega_0<1$;  if $ \nu / \omega_0>1$, it represents that an ion is excited from its ground state by annihilating a phonon but accompanied with the creation of three photons; and  $ \nu / \omega_0<0$ corresponds to the situation discussed in the main body of the text (the \textbf{case (\emph{c})} in the following).

\begin{figure}[btp]
\renewcommand\thefigure{S4}
\centering
\includegraphics[scale=0.38]{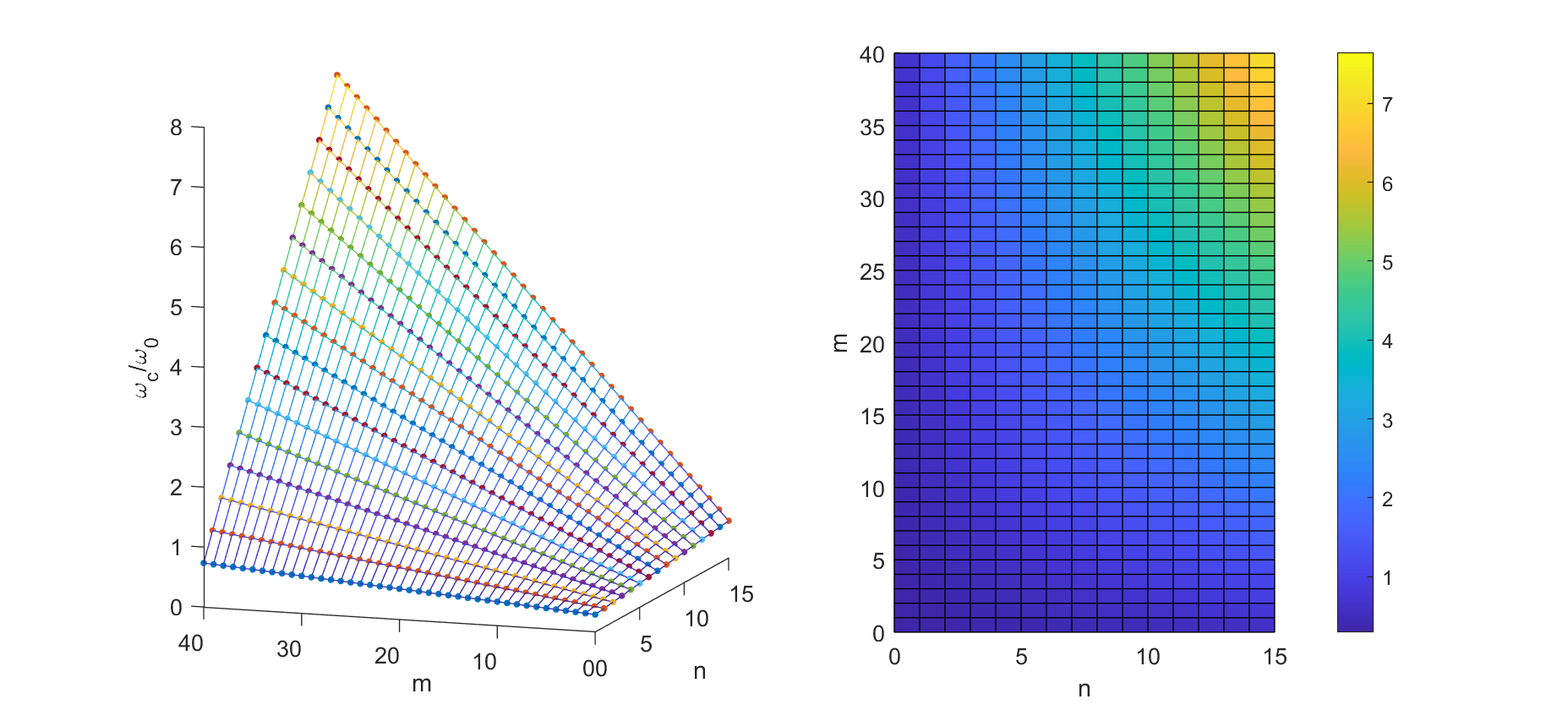}
\caption{Plots of the position of the resonance as functions of photon number $ m $ and phonon number $ n $.
}
\label{Omegc}
\end{figure}

In \textbf{case (\emph{b})}: one ion can be excited by annihilating three photons and three phonons, we find that only one path, which is shown in Fig. \ref{p311a}(c), can connect the states $ \left\vert g,3,3\right\rangle $ and  $ \left\vert e,0,0\right\rangle $.  By taking $ \omega_{c}=(\omega_{0}-3\delta)/3$, it is easy to get the effective coupling rate for the transition between the states $ \left\vert g,3,3\right\rangle $ and $ \left\vert e,0,0\right\rangle$, which is  $  \Omega_{\mathrm{eff}}^{(3b)} =\frac{27(\eta g)^{3}}{2\omega_{0}^{2}}$ \cite{Veff3b}.  
There are two avoided-level crossings in Fig. \ref{Ens311}(b) and the energy splitting in the region around $ \omega_c/\omega_0 \approx 0.18 $ is much smaller than that around $ \omega_c/\omega_0 \approx 0.293 $ if taking the same parameters.

\textbf{Case (\emph{c})}, which is adopted in the main body of the text, is opposite to \textbf{case (\emph{a})} in phonon states. By taking resonance condition $\omega_{c}=(\omega_{0}+\delta)/3$, we find 
\begin{align}
H^{(2c)}_{\mathrm{eff}}&=V^{(2c_1)}_{\mathrm{eff}} \sigma_+\sigma_- -V^{(2c_2)}_{\mathrm{eff}}\sigma_-\sigma_+,\label{H2c}\tag{S8a}\\
H^{(3c)}_{\mathrm{eff}}&=-V_{\mathrm{eff}}^{(3c)}[a^{\dagger}aa^{\dagger}b^{3}\sigma_{+}+aa^{\dagger}a(b^{\dagger})^{3}\sigma_{-}],\label{H3c}\tag{S8b}
\end{align}
where $
V^{(2c_1)}_{\mathrm{eff}} =\dfrac{3(\eta g)^{2}}{2} \left[\dfrac{(n+1)(m+1)}{\omega_0-2\nu} +\dfrac{(n+1)m}{2\omega_0-\nu}+\dfrac{n(3m+2)}{2(\omega_0+\nu)} \right] $, $V^{(2c_2)}_{\mathrm{eff}}=\dfrac{3(\eta g)^{2}}{2} \left[\dfrac{nm}{\omega_0-2\nu}+  \dfrac{n(m+1)}{2\omega_0-\nu} \right.$  $ \left. +\dfrac{(n+1)(3m+1)}{2(\omega_0+\nu)}\right] $ and $ V_{\mathrm{eff}}^{(3c)}=\frac{27\omega_{0}(\eta g)^{3}}{4(\omega_{0}+\nu)^{2}(\omega_{0}-2\nu)}$. 
The effective coupling rate between the states $ \left\vert g,3,0\right\rangle $ and $ \left\vert e,0,1\right\rangle $ is $ \Omega_{\mathrm{eff}}^{(3c)}=\sqrt{6}V_{\mathrm{eff}}^{(3c)} $ \cite{Veff3c}.
After $H_0+H^{(2c)}_{\mathrm{eff}}$ operating on states $ \left\vert g,3,0\right\rangle $ and $ \left\vert e,0,1\right\rangle $,  the position of the resonance is
\begin{equation}\label{Resc}\tag{S9}
\dfrac{\omega'_c}{\omega_0}=\dfrac{1}{3}\left(1+\dfrac{\nu}{\omega_0}\right)+\left(\dfrac{1}{1-2\dfrac{\nu}{\omega_0}}-\dfrac{2}{1+\dfrac{\nu}{\omega_0}} \right)\left(\dfrac{\eta g}{\omega_0}\right)^2 + O\left(\dfrac{\eta g}{\omega_0}\right)^4. 
\end{equation}
Similar to \textbf{case (\emph{a})}, the magnitude of the energy splitting and the position of the resonance shifts with the increase in photon and phonon number.
%When $ \nu/ \omega_{0}=0.5$,  $ V_{\mathrm{eff}}^{(3c)}$  is infinite according to the analytical result. As $ \nu$  approaches to $0.5 \omega_{0}$,  the energy of state  $\left\vert g,1,2\right\rangle  $ gets close to states $\left\vert e,0,1\right\rangle  $ and $\left\vert g,3,0\right\rangle  $, then the intermediate state $\left\vert g,1,2\right\rangle$ may  become a finial state. However, numerical result shows that the energy splitting still exists,  where the divergence derived from the analytical results turn out to be a peak appearing in Fig. \ref{Fig:3DOmega}.  And the results for $ \nu$  approaching to $0.5 \omega_{0}$ coincide with  the period of a complete population oscillation in Sec. IV B 2.
 %as finite photon and phonon number are taken in numerical calculation.  , which is the  . which is confirmed from

When $ \omega_{c}=(\omega_{0}+3\delta)/3$, we get \textbf{case (\emph{d})}: three photons excite one ion accompanied with the creation of three phonons, where the effective Hamiltonian reads
\begin{equation}\label{Heff3b} \tag{S10}
H_{\mathrm{eff}}^{(3d)}=-V_{\mathrm{eff}}^{(3d)}[(a^{\dagger})^{3}b^{3}\sigma_{+}+a^3(b^{\dagger})^{3}\sigma_{-}],
\end{equation}
with $ V_{\mathrm{eff}}^{(3d)}=\frac{9(\eta g)^{3}}{4\omega_{0}^{2}}$.
The effective coupling rate for the transition between the states $ \left\vert g,3,0\right\rangle  $ and $ \left\vert e,0,3\right\rangle$ is $\Omega_{\mathrm{eff}}^{(3d)}= 6V_{\mathrm{eff}}^{(3d)}=\frac{27(\eta g)^{3}}{2\omega_{0}^{2}} $.
Results for \textbf{case (\emph{c})} and \textbf{(\emph{d})} are similar to those for \textbf{case (\emph{a})} and \textbf{(\emph{b})}. \newline

In brief, we have shown that when $ \omega_{c}=(\omega_{0}\pm\delta)/3 $ or $ (\omega_{0}\pm3\delta)/3 $, a resonant three-photon coupling can be constructed. % in this model we proposed here. 
%They can be of interest in the generation of hybrid entangled GHZ-like states, e.g., $\left( \left\vert g,3,1\right\rangle + \left\vert e,0,0\right\rangle \right) /\sqrt{2}$, and the swap of information stored in the trapped ions to the electromagnetic field and vice versa . 
%In the Greenberger-Horne-Zeilinger(GHZ)-like entangled states $\left( \left\vert g,3,0\right\rangle + \left\vert e,0,1\right\rangle \right) /\sqrt{2}$, we get that transfer one phonon stored in the ion to the cavity can create three photons, 
In such a system, GHZ-like entangled states (e.g., $\left( \left\vert g,3,0\right\rangle + \left\vert e,0,1\right\rangle \right) /\sqrt{2}$) can be obtained,  which has potential applications in quantum information storing and quantum communication.

%\newpage
\begin{figure}[btp]
\renewcommand\thefigure{S5}
\centering
\includegraphics[width=4.1 in]{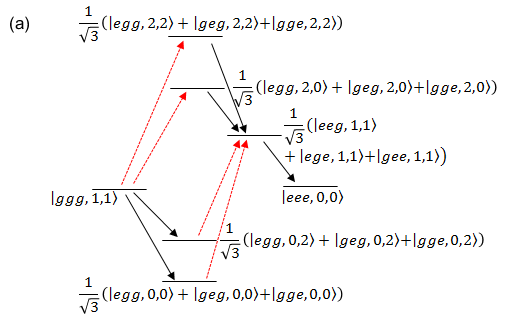}\\
\includegraphics[width=4.1 in]{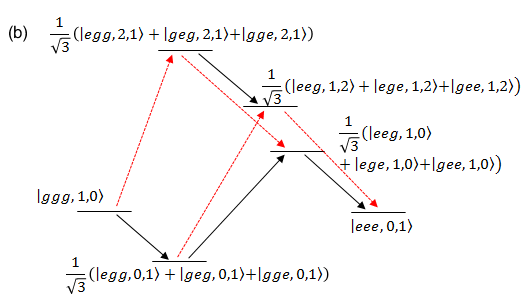}\\
\includegraphics[width=4.1 in]{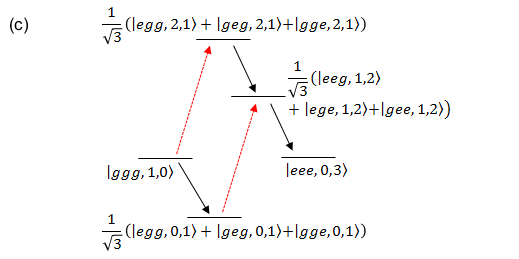}\\
\includegraphics[width=4.1 in]{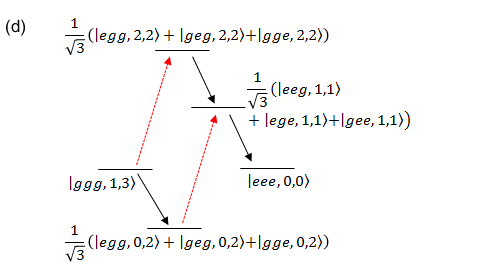}\\
\caption{\label{trans3} Sketches of the high-order transitions for case (\emph{e-h}).} %via intermediate virtual transitions.%, arrowed red dashed line represents the process that excitation and photon-number are nonconserving. 
\end{figure}

%\newpage
\subsubsection{B. Simultaneous excitations of three hot trapped ions with one photon}
To realize the simultaneous excitations of three hot trapped ions with one photon in a cavity, we carefully inspect all possible intermediate states and get the transition paths shown in Fig. \ref{trans3}.
Now we discuss the first \textbf{case (\emph{e})}: one photon and one phonon can be jointly absorbed by the three ions in their ground state which will all reach their excited state. 
To show this we indicate a part of the energy spectrum as a function of $ \omega_c $ in Fig. \ref{3ion} with same parameters as these in Sec. I A, around $ \omega_c/\omega_0\approx 2.8 $ where the simultaneous excitations arise, what appears as a crossing on this scale turns out to be a splitting anticrossing on an enlarged view.
Observing that just outside this avoided-crossing region one level remains flat, while the other grows as $ \omega_c $, this splitting clearly originates from the
hybridization of the states $ \left\vert ggg,1,1\right\rangle $ and $ \left\vert eee,0,0\right\rangle $. The resulting states are well approximated by the states
$\left(  \left\vert ggg,1,1\right\rangle \pm \left\vert eee,0,0\right\rangle \right)  /\sqrt{2}$.

\begin{figure}[bp]
\renewcommand\thefigure{S6}
\centering%\includegraphics[scale=0.66]{Transitiona2.png}
\includegraphics[width=4.3 in]{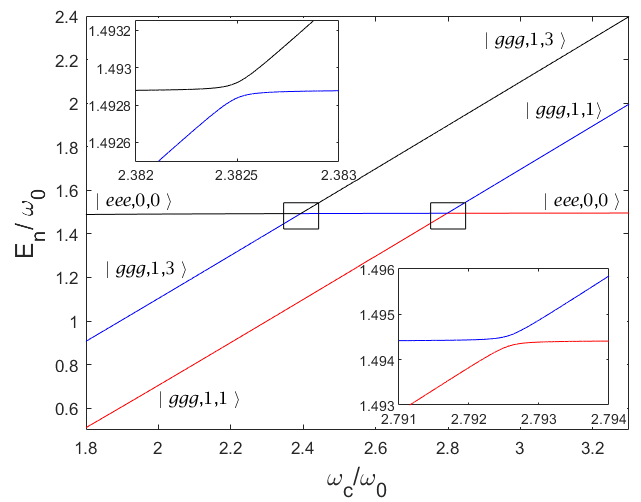}
\caption{%(a) All virtual transitions that contribute to the process $ \left\vert ggg,1,1\right\rangle \rightarrow \left\vert eee,0,0\right\rangle $ for case (\emph{c}) and (b) the process $ \left\vert ggg,1,3\right\rangle \rightarrow \left\vert eee,0,0\right\rangle $ for case (\emph{f}). (c) 
Plot of $ E_n/\omega_0 $  as functions of $ \omega_c/\omega_0 $ with $ \nu/\omega_0 =0.2$, $ \eta g/\omega_0 =0.06$. An avoided-level crossing takes place at $ \omega_c' \approx 2.7925 \omega_0$, demonstrating the coupling between the states $ \left\vert ggg,1,1\right\rangle $ and $ \left\vert eee,0,0\right\rangle $ with the magnitude of the energy splitting about $2.3\times10^{-4} \omega_0 $, and at $ \omega_c'' \approx 2.3825 \omega_0$  the magnitude of the energy splitting between the states $ \left\vert ggg,1,3\right\rangle $ and $ \left\vert eee,0,0\right\rangle $ is $8.3\times10^{-5} \omega_0 $  . 
}
\label{3ion}
\end{figure}

Now we apply the James method \cite{james02} to derive the effective Hamiltonian.  In the interaction picture with respect to $ H_0 $, the Hamiltonian is given by
\begin{equation}  \label{Hi} \tag{S11}
\begin{split}
H_{I}=\eta g \sum_{i=1,2,3}(a^{\dagger}e^{i\nu t}+ae^{-i\nu t})
(b^{\dagger}e^{i\omega_c t}+be^{-i\omega_c t})(\sigma^i_{+}e^{i\omega_0 t}+\sigma^i_{-}e^{-i\omega_0 t}). 
\end{split}
\end{equation}
When $\omega_{c}=3\omega_{0}-\nu $, $ H_I $ expressed in Eq. \eqref{Hi} possesses four distinct frequencies 
$ \omega_1=2\omega_0 -\nu$, $ \omega_2=2\omega_0 $, $ \omega_3=4\omega_0-\nu $ and $ \omega_4=4\omega_0 $ (satisfying $\omega_4=2\omega_2  $ and $ \omega_3=\omega_1+ \omega_2$), the corresponding $ h_m $  $(m = 1, 2, 3, 4)$ are of the form 
\begin{align}
 h_1&=\eta g \sum_{i=1,2,3}a b^{\dagger}\sigma^i_{-},  \quad
  h_2=\eta g \sum_{i=1,2,3}a^{\dagger} b^{\dagger}\sigma^i_{-},  \quad
   h_3=\eta g \sum_{i=1,2,3}a b^{\dagger}\sigma^i_{+},  \quad
    h_4=\eta g \sum_{i=1,2,3}a^{\dagger} b^{\dagger}\sigma^i_{+}.\tag{S12}
\end{align}
By straightforwardly utilizing the formula (7) and (15) in Ref. \cite{james02},  we arrive at
\begin{align}
H_{\mathrm{eff}}^{(2e)}&=\sum^{3}_{i,j=1}V_{\mathrm{eff}}^{(2e_{1})}\sigma_{+}^{i}\sigma_{-}^{j}+V_{\mathrm{eff}}^{(2e_{2})}\sigma_{-}^{i}\sigma_{+}^{j},\label{H2e}\tag{S13a}\\
H_{\mathrm{eff}}^{(3e)}&=-V_{\mathrm{eff}}^{(3e)}[ab\sigma_{+}^{1}\sigma_{+}^{2}\sigma_{+}^{3}+a^{\dagger}b^{\dagger}\sigma_{-}^{1}\sigma_{-}^{2}\sigma_{-}^{3}],\label{H3e}\tag{S13b}
\end{align} 
where $ V_{\mathrm{eff}}^{(2e_{1})}= \dfrac{(\eta g)^{2}}{2} \left[ \dfrac{nm}{2\omega_{0}} - \dfrac{(n+1)(m+1)}{\omega_{0}}+ \dfrac{(n+1)m}{2\omega_{0}-\nu}  -\dfrac{n(m+1)}{\omega_{0}-\nu}\right] $, $ V_{\mathrm{eff}}^{(2e_{2})}=  \dfrac{(\eta g)^{2}}{2} \left[ -\dfrac{(n+1)(m+1)}{2\omega_{0}} + \dfrac{nm}{\omega_{0}} \right. $ $ \left. - \dfrac{n(m+1)}{2\omega_{0}-\nu}+\dfrac{(n+1)m}{\omega_{0}-\nu}\right]$ and $ V_{\mathrm{eff}}^{(3e)}=\frac{3\nu(3\omega_{0}-\nu)(\eta g)^{3}}{2\omega_{0}^{2}(\omega_{0}-\nu)(2\omega_{0}-\nu)}$ \cite{three2}. 
%The effective coupling rate $ \Omega_{\mathrm{eff}}^{(3e)} $ can also be obtained by using third-order perturbation theory \cite{Sup2}. 
%One can find that there are only four paths connecting the states $ \left\vert ggg,1,1\right\rangle $ and $ \left\vert eee,0,0\right\rangle $ by carefully inspecting all the possible intermediate states.
Here $ ab\sigma_{+}^{1}\sigma_{+}^{2}\sigma_{+}^{3} $ describes the annihilation of a photon in the cavity accompanied by three ionic excitations from their ground state  with annihilating a phonon, which will take no effect for the vacuum state of cavity mode or vibration mode, and $ a^{\dagger}b^{\dagger}\sigma_{-}^{1}\sigma_{-}^{2}\sigma_{-}^{3} $ describes the reverse process. 
\begin{figure}[btp]
\renewcommand\thefigure{S7}
%\centering
%\includegraphics[width=5.25 in]{Cp01.png}
%\includegraphics[width=4.85 in]{Cp3Done13.png}
%\includegraphics[width=4.250 in]{Comp3D01.png}
\includegraphics[width=5.5 in]{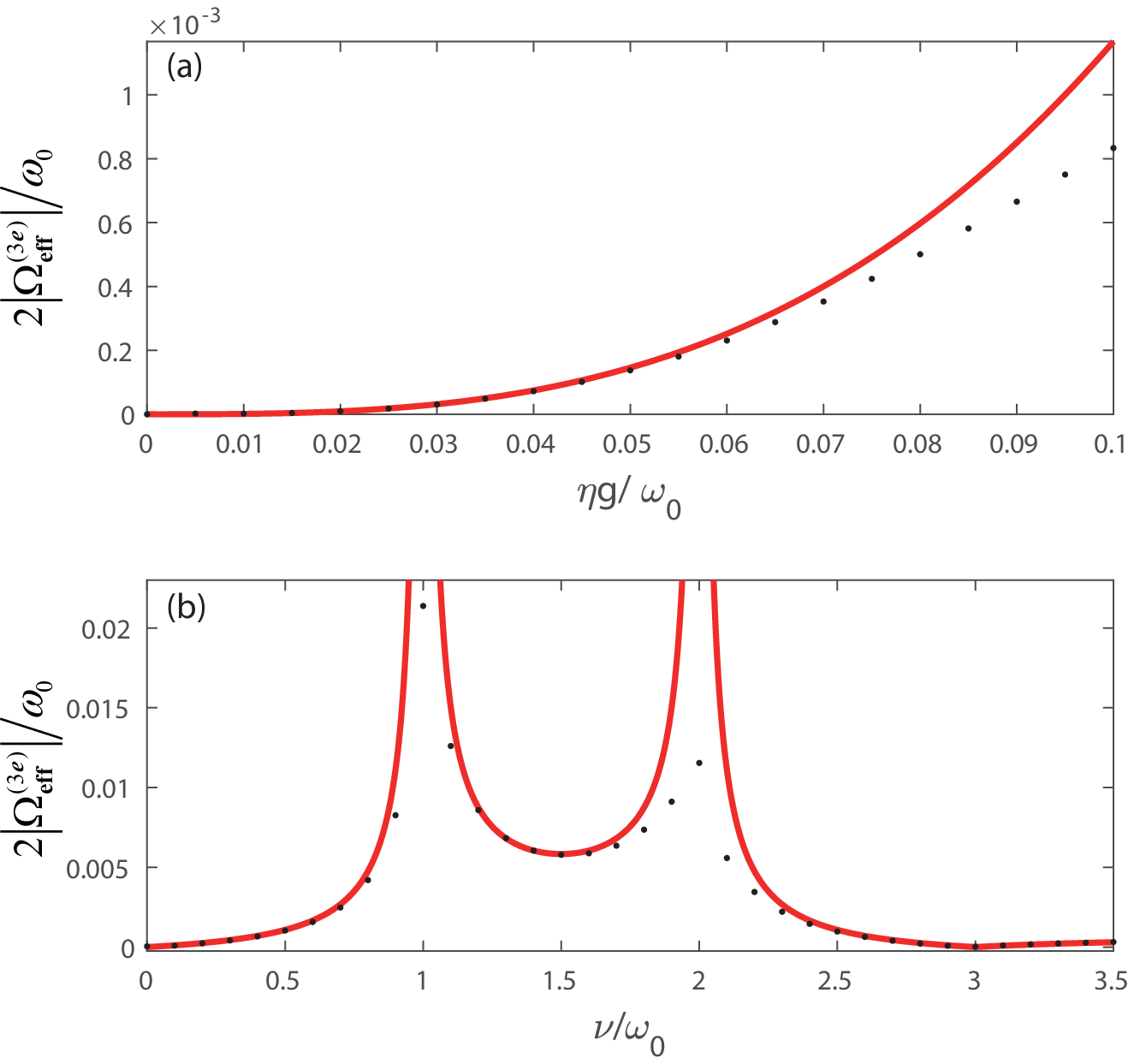}
\caption{(a) Comparison of the magnitudes of the energy splitting $ 2\Omega^{(3e)}_{\mathrm{eff}}/ \omega_0 $ obtained analytically (solid red line) and numerically (black points) as a function of the interaction strength  $ \eta g/ \omega_0$ with $ \nu/\omega_0 =0.2$. (b)  Plot of $2 \vert \Omega_{\mathrm{eff}}^{(3e)} \vert/\omega_0$ in red line as a function  of $ \nu/\omega_0 $ and the energy splitting obtained numerically in black points on resonance  $\omega_{c}+\nu= 3\omega_{0}$ with $ \eta g/\omega_0 =0.06$.
}
\label{Cp113}
\end{figure} 
The effective coupling rate $ \Omega_{\mathrm{eff}}^{(3e)}= V_{\mathrm{eff}}^{(3e)}$ describing the transition between states $ \left\vert ggg,1,1\right\rangle $ and $ \left\vert eee,0,0\right\rangle $ will never go to zero on resonance, which is different to the simultaneous excitation of three atoms with one photon in the appendix of Ref. \cite{Nori02} or the case IV B 2c in Ref. \cite{exp102}  (see details in Sec. III). %\ref{sec:Comparision}  
To check the validity of $ \Omega_{\mathrm{eff}}^{(3e)} $, we compare it with the energy splitting obtained numerically. Fig. \ref{Cp113}(a) presents the results as a function of $ \eta g/ \omega_0$ when $ \nu/\omega_0 =0.2$. $ \Omega_{\mathrm{eff}}^{(3e)} $ agrees well with the numerical results, the percentage difference is less than $5\% $ for $\eta g /\omega_0 <0.045$. 
Fig. \ref{Cp113}(b) displays the comparison of the absolute value of $ 2\Omega_{\mathrm{eff}}^{(3e)} $ obtained analytically and numerically as a function of $ \nu/ \omega_0 $ on resonance $\omega_{c}+\nu= 3\omega_{0}$ when $ \eta g/\omega_0 =0.06$. 
When $ \nu/\omega_0 =1$ or $2$, the value of $ 2 \vert \Omega_{\mathrm{eff}}^{(3e)} \vert$ shows a divergence, which is conformed by numerical calculation with a pronounced but finite peak at the same $\nu/\omega_{0} $. 
Then it is useful to take a fidelity measure near the divergence points.
Besides, when $ \nu $ close to $ 0 $ and $ 3 \omega_0$, $ \Omega_{\mathrm{eff}}^{(3e)} $ infinitely approaches $ 0 $. It means the Rabi splitting vanishes.
Moreover, Eq. (S11) describes the situation $ 0<\nu / \omega_0<3$;  if $ \nu / \omega_0>3$, it means that three ions are simultaneously excited from its ground state by annihilating a phonon but accompanied with the creation of one photon; $ \nu / \omega_0<0$ corresponds to the following \textbf{case (\emph{g})}: one photon  can simultaneously excite three ions and one phonon.

In addition, the effective Hamiltonian \eqref{H2e} allows us to determine the frequency of the cavity field at which simultaneous excitations take place. 
When $ H_0 + H_{\mathrm{eff}}^{(2e)} $ operates on the states $ \left\vert ggg,1,1\right\rangle $ and $ \left\vert eee,0,0\right\rangle $, a solution of the cavity field frequency is given by
\begin{equation}
\dfrac{\omega'_c}{\omega_0}=2.8-\frac{25}{16}\left(\dfrac{\eta g }{\omega_0 } \right)^{2} +O\left(\dfrac{\eta g }{\omega_0 } \right)^{4}, 
\label{Rf}\tag{S14}
\end{equation}
where $\nu /\omega_0= 0.2$ has been taken into consideration. % achieved. % The $(\eta g)^{2}$  dependence results from the Stark shifts of energy levels \cite{Law}. 
Comparing the frequency predicted by Eq. \eqref{Rf} with the numerical results, we find  the percentage difference is less than $ 0.02\% $ for $\eta g /\omega_0 <0.1$.

The second \textbf{case (\emph{f})}: one photon can simultaneously excite three ions with the annihilation of three phonons. By taking $\omega_{c}=3\omega_{0}-3\delta $, the effective coupling strength between the state $ \left\vert ggg,1,3\right\rangle $ and $ \left\vert eee,0,0\right\rangle $ is 
\begin{equation}
\Omega_{\mathrm{eff}}^{(3f)}=\frac{9\sqrt{6}(\nu-\delta)(\eta g)^{3}}{(\omega_{0}-\nu)(2\omega_{0}-3\delta+\nu)(4\omega_{0}-3\delta-\nu)}, \tag{S15} 
\end{equation}
 %of the same order of magnitude. 
which goes to zero when $ \nu=\delta $; that is, the two paths connecting the states $ \left\vert ggg,1,3\right\rangle $ and $ \left\vert eee,0,0\right\rangle $ interfere destructively.
However the coupling, the rate of which may be smaller than the system decay rates and sometimes can be ignored, %is much smaller than $ \Omega_{\mathrm{eff}}^{(3e)} $ seen from if same parameters are taken, 
 evertheless exists as shown in Fig. \ref{3ion}, due to the influence of higher-order processes and the energy levels shifted from their bare-state values to the dressed states \cite{exp102}.  
 
For \textbf{case (\emph{g})}: one photon can simultaneously excite three ions accompanied with the creation of one phonon,  
by utilizing the third-order James' effective Hamiltonian method and taking $\omega_{c}=3\omega_{0}+\nu $ into consideration, we find 
\begin{equation}
H_{\mathrm{eff}}^{(3g)}=-V_{\mathrm{eff}}^{(3g)}[a^{\dagger}b\sigma_{+}^{1}\sigma_{+}^{2}\sigma_{+}^{3}+ab^{\dagger}\sigma_{-}^{1}\sigma_{-}^{2}\sigma_{-}^{3}], \label{Heff3g} \tag{S16} 
\end{equation}
with $ V_{\mathrm{eff}}^{(3g)}=\Omega_{\mathrm{eff}}^{(3g)}=\frac{3\nu(3\omega_{0}+\nu)(\eta g)^{3}}{2\omega_{0}^{2}(\omega_{0}+\nu)(2\omega_{0}+\nu)}$ \cite{Omega3g} being the effective coupling rate for the transition between the states $ \left\vert ggg,1,0\right\rangle $ and $ \left\vert eee,0,1\right\rangle $.

In \textbf{case (\emph{h})}: one photon can simultaneously excite three ions and three phonons, when $\omega_{c}=3\omega_{0}+3\delta $, the effective coupling strength between the state $ \left\vert ggg,1,0\right\rangle $ and $ \left\vert eee,0,3\right\rangle $ is $ \Omega_{\mathrm{eff}}^{(3h)}=\frac{9\sqrt{6}(\delta-\nu)(\eta g)^{3}}{(\omega_{0}+\nu)(2\omega_{0}+3\delta-\nu)(4\omega_{0}+3\delta+\nu)}$, also yielding zero when $ \nu=\delta $. 
Similar results for \textbf{case (\emph{g,h})} can also be achieved. \newline

Anyway, we have shown that one photon can simultaneously excite three hot trapped ions accompanied with the creation or annihilation of one or more phonons. 
%From this scheme, a construction of three-body interaction $ \sigma_{+}^{1}\sigma_{+}^{2}\sigma_{+}^{3}$ and $\sigma_{-}^{1}\sigma_{-}^{2}\sigma_{-}^{3}$ Hamiltonian can be obtained if concentrating only on the ionic internal states without considering photon and phonon mode.%couples well separate trapped ions together, where ions do not have ion-ion interaction \cite{ions2}. %without interaction between ions 
This scheme can be use to realize three-body effective interaction among trapped ions and to generate complex entangled states, which give a new impulse to develop novel experimental techniques to generate and manipulate quantum states.
%explore multi-qubit physics and to

\clearpage
 
\subsection*{III. Comparison of  the effective coupling rate}\label{sec:Comparision}

The simultaneous excitation of three atoms with a single photon in the supplemental material of Ref. \cite{Nori02} or the case IV B 2c of Ref. \cite{exp102} is to implement the process $ \left\vert eee,0\right\rangle \leftrightarrow \left\vert ggg,1\right\rangle $.  As presented in Ref. \cite{exp102}, the effective coupling strength between the states $ \left\vert eee,0\right\rangle $ and $ \left\vert ggg,1\right\rangle $ is
\begin{equation}\label{geff}\tag{S17}
\Omega_{\mathrm{eff}}=  \dfrac{- 3g^3(\omega_c-3\omega_0)}{\omega_0(\omega_c-\omega_0)^2},
\end{equation}
which  \textit{goes to zero} on resonance ($ \omega_c=3\omega_0 $). However, a coupling between the states $ \left\vert eee,0\right\rangle $ and $ \left\vert ggg,1\right\rangle $ nevertheless exists close to that resonance, which is shown through numerical calculations in Sec. II on the supplemental material of Ref. \cite{Nori02} when the two states are \textit{slightly out of resonance}. This is due to the influence of higher-order processes and the energy levels shifted from their bare-state values to the dressed states \cite{exp102}. \newline
%The absolute value of $ g_{\mathrm{eff}} $ is displayed in Fig. \ref{Ocompar}(a) as a function of $ \omega_c $, which has a minimum value approaching to zero at resonant point.

However, when there exists a vibrational mode in the process of simultaneously exciting three atoms with a single photon, the effective coupling strength will \textit{never go to zero} on resonance. In \textbf{case (\emph{e})}: three ions can be  simultaneously excited by annihilating one photon and one phonon, the effective coupling strength for the transition between the states  $ \left\vert eee,0,0\right\rangle $ and $ \left\vert ggg,1,1\right\rangle $ is
\begin{equation}\label{O3e}\tag{S18}
\Omega_{\mathrm{eff}}^{(3e)}= \dfrac{6(\eta g)^3}{\omega_c-\omega_0-\nu}\left(\dfrac{1}{2\omega_0}+\dfrac{1}{\omega_0-\nu}-\dfrac{1}{\omega_c-\omega_0}-\dfrac{2}{\omega_c-\omega_0+\nu} \right).
\end{equation} 
which will \textit{never go to zero} on resonance ($ \omega_c=3\omega_0 -\nu$).
When $ \omega_c=3\omega_0 -\nu$, Eq. \eqref{O3e} just is the $\Omega_{\mathrm{eff}}^{(3e)}  $ in Eq. \eqref{H3e}, then we can directly get the nonzero conclusion.
%Fig. \ref{Ocompar}(b)  displays the absolute value of $ \Omega_{\mathrm{eff}}^{(3e)} $ as  functions of $ \omega_c $. Near the resonance $ \omega_c/\omega_0 =2.8$, $ \Omega_{\mathrm{eff}}^{(3e)} $ increases or declines with $ \omega_c $ increasing. which  on resonance ($ \omega_c=3\omega_0 -\nu$).
And for \textbf{case (\emph{g})}: one photon can simultaneously excite three ions accompanied with the creation of one phonon, similar results can be obtained. The effective coupling strength for the transition between the states  $ \left\vert eee,0,1\right\rangle $ and $ \left\vert ggg,1,0\right\rangle $ is
\begin{align}
\Omega_{\mathrm{eff}}^{(3g)}=&3(\eta g)^3  \left(\dfrac{1}{\omega_0}+\dfrac{2}{\omega_0-\omega_c} \right)\left(\dfrac{1}{\omega_0-\omega_c+\nu}+\dfrac{2}{\omega_0-\omega_c-\nu} \right) , \label{O3g}\tag{S19} 
\end{align} 
which will also \textit{never go to zero} on resonance ($ \omega_c=3\omega_0 +\nu$). 
% %%&3(\eta g)^3  \left(\dfrac{1}{\omega_0}+\dfrac{2}{\omega_0+\nu} \right)\left(\dfrac{1}{\omega_0-\omega_c+\nu}+\dfrac{2}{\omega_0+\omega_c+\nu} \right),\label{O3g2}\tag{S14b}

\clearpage

\subsection*{IV. Adiabatic Landau-Zener transitions}

\begin{figure}[bp]
\renewcommand\thefigure{S8}
%\centering
\includegraphics[width=4.8 in]{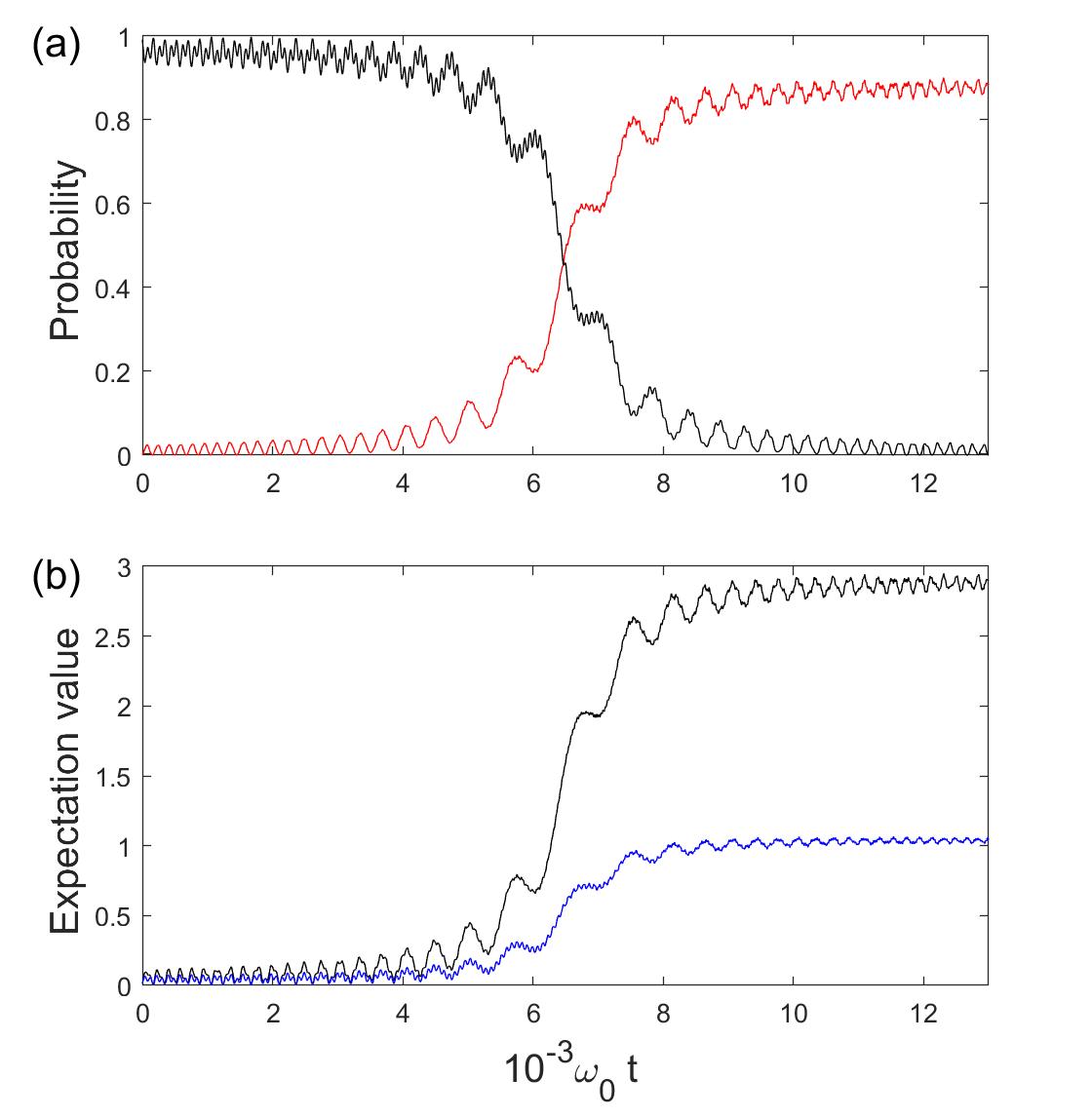}
\caption{(a) The time-dependent Landau-Zener transition process with the states $ \left\vert e,0,0\right\rangle $ (black curve) and $ \left\vert g,3,1\right\rangle $ (red curve). % in the up row. 
(b) Expectation value of the photon number $ \langle b^{\dagger}b\rangle $ (black curve) and phonon number $ \langle a^{\dagger}a\rangle $ (blue curve). % in the down row. 
All results are plotted as a function of $ \omega_0t $ with
$ \eta g/ \omega_0=0.06 $, $ v=2\times10^{-6}\omega_0^2 $ and $ \omega_c(0)=0.28 \omega_0$. }
\label{LZ311} 
\end{figure}

For \textbf{case (\emph{a})}, we first examine the adiabatic Landau-Zener transition effect \cite{Nori202,Law02} without damping around the anticrossing point. % considering the dissipative channels. 
Assume that the photon frequency $ \omega_c $ is linearly dependent in time
\begin{equation}
\omega_c(t) = \omega_c(0) + vt, \quad 0 \leq t \leq t_f , \label{freq}\tag{S20} 
\end{equation}
such that $ \omega_c(t) $  can sweep across the resonance [i.e., $ \omega_c(t') = \omega_c'$] at a certain time $t' < t_f$. The sweeping speed  $ v $ controls the adiabaticity of the process \cite{Law02}. 
Initially, the system is prepared in its $ j $th eigenstate $ \left\vert \psi_j\right\rangle\simeq \left\vert e,0,0\right\rangle $. Due to the diabatic transition, the system evolves far away from a quasisteady state and might jump to the lower eigenstate $ \left\vert \psi_{j-1}\right\rangle$. 
Therefore, we can estimate the final transition probability to the state $ \left\vert \psi_{j-1}\right\rangle\simeq \left\vert e,0,0\right\rangle $, expressed by the Landau-Zener formula
\begin{equation}
P_{LZ } = \exp \left[ -2\pi\dfrac{\Omega^2_{\mathrm{eff}}}{ d\Delta E/dt} \right],\label{Plz} \tag{S21}
\end{equation}
where $ \Delta E $ is the energy difference between the two diabatic states. Note that $ P_{LZ } $ is the probability of diabatic transition, and it should be as small as possible. From the effective Hamiltonian \eqref{H3a} and the linear sweeping in Eq. \eqref{freq}, we simply have $ d\Delta E/dt\approx 3v $.
If the energy-sweeping speed $v$ is extremely slow and satisfies the relation $3v\ll 2\pi \Omega^2_{\mathrm{eff}}$, the anticrossing point traverses adiabatically. In this case, the system approximately evolves along the $ j $th-energy curve, and the system rarely jumps to the $(j-1)$th eigenstate after the sweeping \cite{Nori202}. \newline

By numerically simulating the evolution dominated by the Schr{\"o}dinger equation, we plot the probabilities of the states $ \left\vert g,3,1\right\rangle $ and $ \left\vert e,0,0\right\rangle $ varying with time, respectively.  In Fig. \ref{LZ311}(a), the probability of state $ \left\vert g,3,1\right\rangle $ gradually increases from $ 0 $ to $ \sim 0.9 $. This small deviation is comprehended due to the existence of weak diabatic-transition effect, other states have low but nonzero probabilities. 
The dynamics of this Landau-Zener transition evidences the resonant coupling between the states $ \left\vert g,3,1\right\rangle $ and $ \left\vert e,0,0\right\rangle $ \cite{Nori202}.  
In addition, three-photon generation indicated by the expectation value of the photon number $ \langle b^{\dagger}b\rangle $ is shown in Fig. \ref{LZ311}(b), so is one-phonon generation indicated by the expectation value of the phonon number $ \langle a^{\dagger}a\rangle $. For all the other cases one can get similar results when examining the adiabatic Landau-Zener transition effect.

\clearpage
\subsection*{V. Quantum Rabi oscillations and damping effects}

\begin{figure}[bp]
\renewcommand\thefigure{S9}
\centering
\includegraphics[width=5 in]{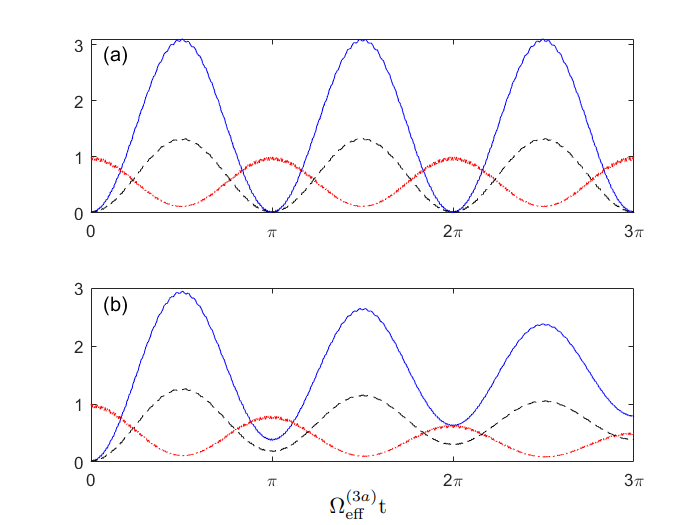}
\caption{Time evolution of the ion mean excitation number $ \langle C^- C^+ \rangle $ (red dot-dashed curve), the cavity mean photon number $ \langle X^- X^+ \rangle $ (blue solid curve) and the vibration mean phonon number $ \langle P^- P^+ \rangle $ (black dotted curve), assuming the system is initially in the ionic excited state $ \left\vert e,0,0\right\rangle $. (a) System dynamics with no decay. (b) %System dynamics with ultra-high heating rate $ \zeta= 1\times10^{-3}\omega_0$, $ \gamma=\kappa=0 $. (c) 
System dynamics with decay rates $ \kappa=\gamma= 2\zeta=1\times10^{-4}\omega_0 $. Other parameters are the same as in Fig. \ref{Ens311}.
}
\label{311TE2}
\end{figure}

To examine the deterministic transition between the states $\left\vert g,3,1\right\rangle $ and $ \left\vert e,0,0\right\rangle $, the rate of the adiabatic Landau-Zener transition process is extremely slow. Therefore, we can simply observe the Rabi oscillation between these two states.
As the distinction between bare (unobservable) excitations and physical particles that can be detected, we therefore calculate the output signals and correlations that can be measured in a photodetection experiment. 
If the resonator, ionic internal state and vibration mode are coupled to the vacuum environment and the initial states of the system are prepared on their ground states  $ \left\vert g,0,0\right\rangle $. In order to probe the anomalous avoided crossing shown in Figs. \ref{Ens311}, we directly excite the ion via a microwave antenna by an optical Gaussian pulse \cite{Lg02}. %In cavity QED systems, this emission can be detected by coupling the ion to an additional microwave antenna through applying an optical Gaussian pulse. 
The corresponding driving Hamiltonian is 
\begin{equation}
H_d=A\exp\left[-(t-t_0)^2/(2\tau^2)\right]/(\tau\sqrt{2\pi})\cos(\omega t)(\sigma^i_- + \sigma^i_+),\label{Hd} \tag{S22}
\end{equation}  
where $ A $ and $ \tau $ are the amplitude and the standard deviation of the Gaussian pulse, respectively. The central frequency of the pulse takes $ \omega= (E_{e00}+E_{g31})/\hbar$. Thus the total Hamiltonian of the system under the influence is $ H_{total}=H+H_d $. The emission field for the $ i $th ion is proportional to zero-time delay correlation function %the ion mean excitation number 
$ \langle C^- C^+ \rangle $, where $ C^{\pm} $ are the ion positive and negative frequency operators, defined as $ C_i^+=\sum_{j,k>j}C^i_{jk}  \vert j\rangle \langle k \vert$ and $C^-= (C^+)^{\dagger} $, with $ C^i_{jk}\equiv  \langle j \vert (\sigma^i_- + \sigma^i_+) \vert k\rangle$. The output photon flux emitted by a resonator can be expressed as $ \Phi_{out} = \kappa \langle X^- X^+ \rangle $, where $ X^+=\sum_{j,k>j}X_{jk}  \vert j\rangle \langle k \vert$ and $X^-= ( X^+)^{\dagger} $ are the positive and negative frequency cavity-photon operators, with $ X_{jk}\equiv  \langle j \vert (b+b^{\dagger}) \vert k\rangle$. Similarly, the output phonon flux is proportional to the the mean phonon number $ \langle P^- P^+ \rangle $, where $ P^+=\sum_{j,k>j}P_{jk}  \vert j\rangle \langle k \vert$ and $P^-= (P^+)^{\dagger} $, with $ P_{jk}\equiv  \langle j \vert (a+a^{\dagger}) \vert k\rangle$.  Neglecting the CRTs, or in the limit of negligible coupling rates, $ X^+ $, $ P^+ $ and $ C^+ $ coincide with $ b $, $ a $ and $ \sigma_- $, respectively \cite{Nori02, Yin02}. 

\subsubsection*{A. System dynamics for \textbf{case (a)}: three photons and one phonon excite  one ion  \cite{Nori202,Yin02, Lg02}}

Firstly, we assume that the ion has been excited and the system is prepared in the initial state $ \left\vert e,0,0\right\rangle $, the time evolution of the cavity mean photon number $ \langle X^- X^+ \rangle $, the ion mean excitation number $ \langle C^- C^+ \rangle $ and the vibration mean phonon number $ \langle P^- P^+ \rangle $ is shown in Fig. \ref{311TE2}. 
Fig. \ref{311TE2}(a) display the ideal case that all the decay rates are zero. %the ion mean excitation number varies cosinoidally and meanwhile, the mean photon and phonon number varies sinusoidally with the same frequency.
In such a case, the mean photon number at its maximum is very close to $ 3 $ , which is an signature that the cavity mode is in a three-photon state. Meanwhile,  the ion jumps to its ground state and the mean phonon number close to $ 1 $ gets its maximum. 
This process is reversible and reveals an energy exchange. 
%We observe that the peak values of $ G^{(3)}$ are approximately two times higher than those of the mean photon number $ \langle X^- X^+ \rangle $, indicating an almost-perfect three-photon correlation.
Fig. \ref{311TE2}(b) shows the system dynamics including loss effects, the mean values still oscillate in cosine or sine form, but their amplitudes as expected decrease exponentially.  \newline
\begin{figure}[bp]
\centering
\renewcommand\thefigure{S10}
\includegraphics[width=5 in]{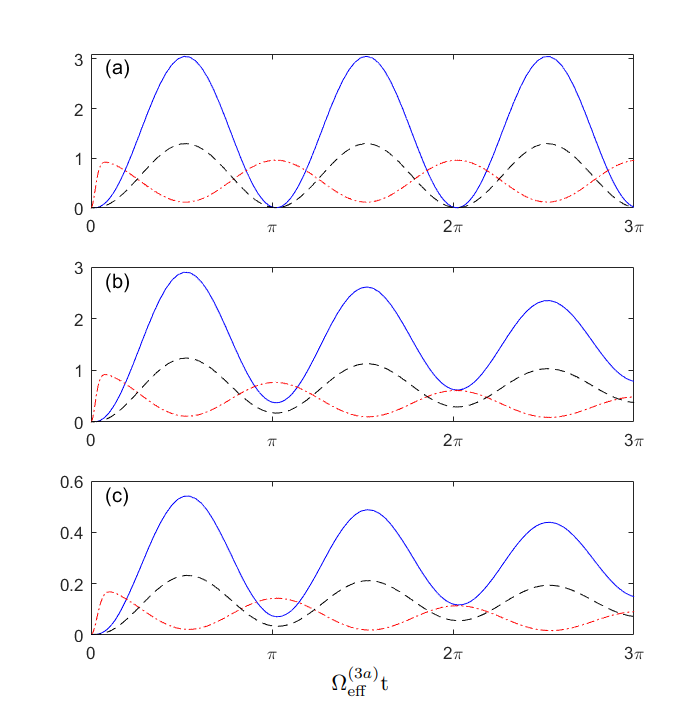}
\caption{Time evolution of the cavity mean photon number $ \langle X^- X^+ \rangle $ (blue solid curve), the ion mean excitation number $ \langle C^- C^+ \rangle $ (red dot-dashed curve), and the vibration mean phonon number $ \langle P^- P^+ \rangle $ (dashed black curve) after the arrival of a $ \pi $-like Gaussian pulse initially exciting the ion. (a) System dynamics without damping effect when  $ A=g $. (b) System dynamics with $ A=g $. (c) System dynamics with $ A=0.3g $. The decay rates for (b,c) are $ \gamma=\kappa=2 \zeta=1\times10^{-4}\omega_0$. Other parameters are the same as in Fig. \ref{Ens311}.
}
\label{311drive}
\end{figure}

\begin{figure}[btp]
\renewcommand\thefigure{S11}
%\centering
\includegraphics[width=5 in]{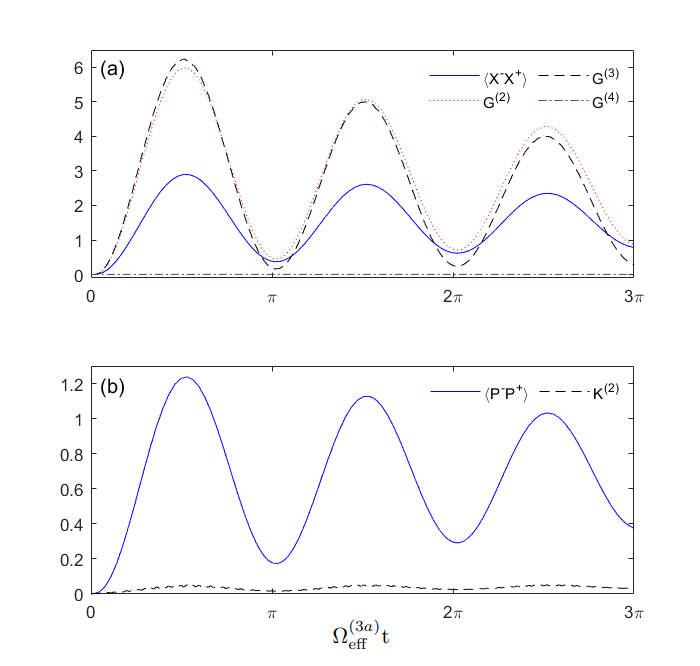}
\caption{(a) Time evolution of the zero-delay three-photon function $ G^{(3)} $ (dashed black curve), two-photon function $ G^{(2)} $ (dotted red curve) and four-photon function $ G^{(4)} $ (dot-dashed violet curve) together with the cavity mean photon number $ \langle X^-X^+\rangle $ (solid blue curve). The first peak value of the three-photon correlation function is approximately two times higher than that of the mean photon number, a signature of an almost-perfect three-photon correlation. (b) Time evolution of the mean phonon number $ \langle P^-P^+\rangle $ (solid blue curve) and the zero-delay two-phonon function $ K^{(2)} $ (dashed black curve).  The parameters are the same as in Fig. \ref{311drive}(b).
}
\label{311G3}
\end{figure} 

In order to simulate a more realistic scene, we prepare the system in its ground state $ \left\vert g,0,0\right\rangle $ and  excite the ion via a microwave antenna with a $ \pi $-like Gaussian pulse. %After the arrival of the pulse, the system undergoes vacuum Rabi oscillations, which shows this effective coupling is coherent and reversible. 
Fig. \ref{311drive}(a) displays the numerically calculated dynamics of the mean photon number $ \langle X^- X^+ \rangle $, of the mean excitation number  $ \langle C^- C^+ \rangle $ and of the mean phonon number $ \langle P^- P^+ \rangle $. 
After the arrival of the pulse, the system undergoes vacuum Rabi oscillations showing the reversible excitation exchange. 
It is clearly shown that the mean excitation number $ \langle C^- C^+ \rangle $ grows rapidly when the pulse comes. After $ \langle C^- C^+ \rangle $ getting its maximum, the mean photon number $ \langle X^- X^+ \rangle $ and mean phonon number $ \langle P^- P^+ \rangle $ starts to grow and gets its maximum when $ \langle C^- C^+ \rangle $ gets its minimum.
In Fig. \ref{311drive}(b,c), the mean values oscillate in cosine or sine form with their amplitudes decreasing exponentially when considering loss effects, and the amplitude $ A $ only affects the amplitude value of the three quantities.  
%Here three-photon plays main role in the excitation of one ion, such a phenomenon can also occur with annihilating three phonons. Exactly speaking, the phonon energy should be much smaller than photon energy and ionic transition energy, if so the results will be more accurate. 
Fig. \ref{311G3} displays the time evolution of the zero-delay three-photon correlation function $ G^{(3)}(t)=  \langle X^-(t)  X^-(t) X^-(t) X^+(t) X^+(t)X^+(t) \rangle$, two-photon function $ G^{(2)}=\langle X^-(t)  X^-(t) X^+(t)X^+(t) \rangle $  and four-photon function $ G^{(4)} =\langle X^-(t)  X^-(t) X^-(t) X^-(t) X^+(t) X^+(t) X^+(t)X^+(t) \rangle$ together with the cavity mean photon number $ \langle X^-X^+\rangle $, also the time evolution of the zero-delay two-phonon function $ K^{(2)} =\langle P^-(t)P^-(t)P^+(t)P^+(t)\rangle$ and  the mean phonon number $ \langle P^-P^+\rangle $ are shown here. 
%three-phonon function $ K^{(3)}=\langle P^-(t)P^-(t)P^-(t)P^+(t)P^+(t)P^+(t)\rangle$ together with
The peak values of $ G^{(3)} $ in the first oscillation cycle are approximately twice higher than those of $ \langle X^-X^+\rangle $, indicating a good three-photon correlation \cite{Lg02}. 
And  $ G^{(3)} $ and  $ G^{(2)} $ almost coincide, which indicates the probability of the system to emit three photons is equal to the probability to emit two photons.
During the evolution, higher values of the minima $ \langle X^-X^+\rangle $ and $ \langle P^-P^+\rangle $  is caused by the photon and phonon escape.
However, the almost zero three-photon correlation function $ G^{(3)} $ indicates the ion absorbs three photons every time. 
At early times, $ G^{(3)} $ reaches a peak value slightly beyond $ 6 $ and $ \langle P^-P^+\rangle $ reaches a peak value slightly beyond $ 1 $, it indicates that the system has a probability to emit more than three photons and one phonon, which is  confirmed by the nonzero four-photon correlation function $ G^{(4)} $ and nonzero two-phonon function $ K^{(2)} $. %This small contribution cannot be observed if its low frequency is outside the frequency-detection window.
When the three-photon correlation function $ G^{(3)}(t)$ reaches its highest value. Meanwhile, the mean excitation number $ \langle C^- C^+ \rangle $ reaches it lowest value and converts into the excitations of three photons and one phonon. The reversible evolution is due to the vacuum Rabi oscillations between the states $\left\vert g,3,1\right\rangle $ and $ \left\vert e,0,0\right\rangle $.

%\begin{figure}[bp]
%\renewcommand\thefigure{S9}%\centering
%\includegraphics[width=6.5 in]{311damp.png}
%\caption{Effects of heating rate (dashed curves) on the dynamics of (a) the mean photon number  $ \langle X^-X^+\rangle $, (b) the effective ion population  $ \langle C^-C^+\rangle $, (c) the three-photon correlation function $ G^{(3)} $, and (d) the mean phonon number  $ \langle P^-P^+\rangle $. Calculations have been performed with the same parameters as in Fig. \ref{311drive}(b). Solid curves display numerical results obtained in the absence of damping for vibration mode, $ \zeta=0$.}
%\label{311damp}
%\end{figure} 

% Fig. \ref{311damp} shows the dynamics of $ \langle X^-X^+\rangle $, $ \langle C^-C^+\rangle $, $ G^{(3)} $ and $ \langle P^-P^+\rangle $ in the absence and existence of damping for vibration mode.

\subsubsection*{B. Time evolution for \textbf{case (c)}: three photons excite one ion and one phonon}

When $ \nu /\omega_0 $  approaches to $ 1/2 $, the numerically-calculated  dynamics of the system exhibit a two-frequency Rabi oscillation shown in Fig. \ref{two-freq}. As $ \nu $ gets close to $ \omega_0/2 $, the energy of state  $\left\vert g,1,2\right\rangle  $ closes to that of states $\left\vert e,0,1\right\rangle  $ and $\left\vert g,3,0\right\rangle  $, then the intermediate state $\left\vert g,1,2\right\rangle  $ becomes a finial state. This induces a three-state resonance, i.e., $\left\vert g,1,2\right\rangle  $,  $\left\vert e,0,1\right\rangle  $ and $\left\vert g,3,0\right\rangle  $.
%  Besides, the results show a little bit difference if the initial state of the system is $\left\vert e,0,1\right\rangle  $ and  $\left\vert g,3,0\right\rangle $, respectively. 
Fig. \ref{two-freq} displays the time evolution of the cavity mean photon number $ \langle X^- X^+ \rangle $, the ion  mean excitation number $ \langle C^- C^+ \rangle $ and the mean phonon number $ \langle P^- P^+ \rangle $ when the initial state of the system is prepared as $\left\vert e,0,1\right\rangle  $ and  $\left\vert g,3,0\right\rangle $,  respectively. 
The mean values present a similar periodic oscillation if starting from different initial state, 
while when preparing the initial state as $\left\vert e,0,1\right\rangle $, the values also oscillate rapidly in the period of a complete population oscillation, which obviously exhibit a two-frequency Rabi oscillation. 
By applying perturbation theory, one can find that the initial state $\left\vert g,3,0\right\rangle  $ via  a intermediate state $\left\vert e,2,1\right\rangle  $ can  reach  the final state  $\left\vert g,1,2\right\rangle $, and the effective coupling strength between states $\left\vert g,3,0\right\rangle  $ and $\left\vert g,1,2\right\rangle $  is close to $ \Omega^{(3c)}_{\mathrm{eff}} $. However,  the effective coupling strength between states  $\left\vert e,0,1\right\rangle  $ and $\left\vert g,1,2\right\rangle $ is much large than $ \Omega^{(3c)}_{\mathrm{eff}} $ as the initial state $\left\vert e,0,1\right\rangle  $ can directly reach the final state $\left\vert g,1,2\right\rangle  $ without intermediate state. 
Higher-order processes can also contribute, e.g., $\left\vert e,0,1\right\rangle \longrightarrow \left\vert g,1,0\right\rangle \longrightarrow  \left\vert e,2,1\right\rangle \longrightarrow \left\vert g,1,2\right\rangle $, but giving little contribution. 
When $ \nu $ gets closer to $ \omega_0/2 $, the two-frequency Rabi oscillation shows more obviously, especially  when the initial state of the system is  $\left\vert e,0,1\right\rangle  $.

\begin{figure}[btp]
\renewcommand\thefigure{S12}
\includegraphics[width=7.65 in]{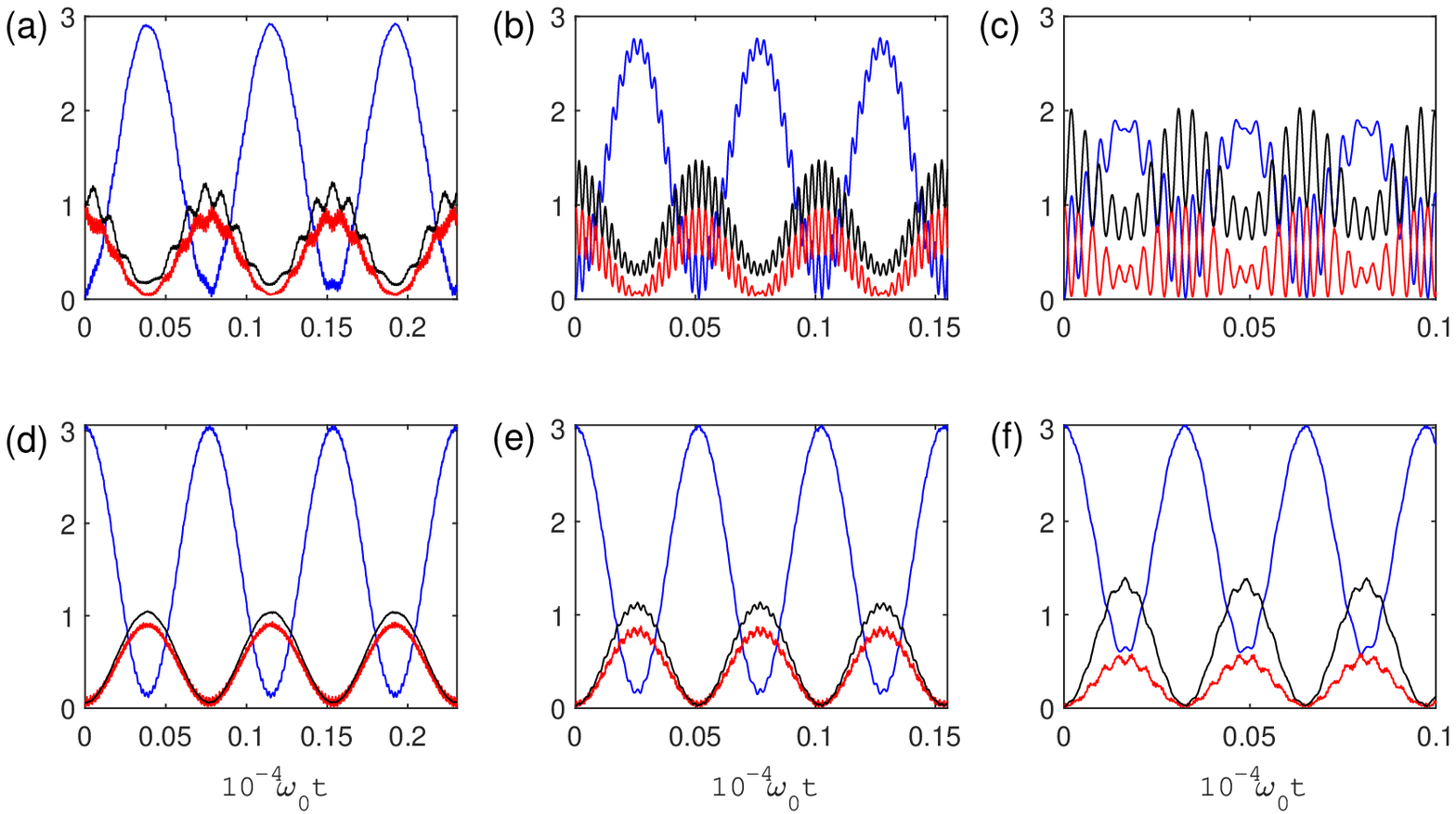}
\caption{Time evolution of the cavity mean photon number $ \langle X^- X^+ \rangle $ (blue  curve), the ion  mean excitation number $ \langle C^- C^+ \rangle $ (red curve) and the mean phonon number $ \langle P^- P^+ \rangle $ (black curve) by preparing the initial state of the system as $\left\vert e,0,1\right\rangle $ for (a,b,c)  and $\left\vert g,3,0\right\rangle  $ for (d,e,f), with  parameters (a,d) $ \nu=0.2\omega_0 $, (b,e) $ \nu=0.4\omega_0 $ and (c,f) $ \nu=0.5\omega_0  $.}
\label{two-freq}
\end{figure}

%
%
%\begin{figure}[bp]
%\renewcommand\thefigure{S13}
%\centering
%\includegraphics[width=7.25 in]{311TEC45and5.png}
%\caption{(a,c,e,g) display the time evolution of the cavity mean photon number $ \langle X^- X^+ \rangle $ (blue  curve), the ion  mean excitation number $ \langle C^- C^+ \rangle $ (red curve) and the mean phonon number $ \langle P^- P^+ \rangle $ (black curve), (b,d,f,h) show the time evolution of the probability  for states $\left\vert e,0,1\right\rangle  $ (red   curve),  $\left\vert g,3,0\right\rangle  $ (black   curve) and $\left\vert g,1,2\right\rangle  $ (blue  curve).  
%The initial state of the system is $\left\vert e,0,1\right\rangle  $ for (a,b,e,f), and  $\left\vert g,3,0\right\rangle  $ for (c,d,g,h), with $ \nu=0.45 $ for (a-d)  and $ \nu=0.5\omega_0  $ for (e-h).}
%\label{311TE}
%\end{figure}

\subsubsection*{C. System dynamics for \textbf{case (e)}: three-ion excitation with one photon and one phonon \cite{Nori02}}

A small energy splitting means a long periodic time, where such a long time is hard to achieve in realistic experiment.  
And if the effective coupling rate is close to and sometimes even smaller than the system decay rates, the amplitudes of the mean values will decrease rapidly within one period of the population oscillation when including loss effects.

Fig. \ref{oneTE15}  displays the time evolution of the cavity mean photon number $ \langle X^- X^+ \rangle $, mean excitation number $ \langle C_1^- C_1^+ \rangle $ for ion 1 (coincides with that of ion 2 and 3), the two-ion correlation $ S^{(2)}=\langle C_1^- C_2^- C_2^+C_1^+ \rangle $ (coincides with $ \langle C_1^- C_3^- C_3^+C_1^+ \rangle $ and $  \langle C_2^- C_3^- C_3^+C_2^+ \rangle $), the three-ion correlation $ S^{(3)}=\langle C_1^- C_2^-C_3^- C_3^+ C_2^+C_1^+ \rangle $ and the mean phonon number $ \langle P^- P^+ \rangle $ by preparing all ions in their excited states. 
In an ideal case without decay, the mean photon number at its maximum and the ion jumps to its ground state. Meanwhile, the mean phonon number close to 1 gets its maximum. 
The Rabi oscillations show the reversible excitation exchange. 
We observe that the single-ion excitation  $ \langle C_i^- C_i^+ \rangle $  and   $ S^{(2)}$, $ S^{(3)} $  almost coincide at any time. This almost-perfect three-ion correlation is a clear signature of the joint excitation.
When taking loss effects into consideration, the mean values, which decrease exponentially as expected, still oscillate in cosine or sine form. 
As expected, the three-ion correlation is more fragile to losses than two-ion correlation, which are all more fragile to loss than $ \langle C_i^- C_i^+ \rangle $.

To simulate a more realistic scene, one needs to excite three ions to get initial state  $ \left\vert eee,0,0\right\rangle $, or inject both a single photon and a single phonon to prepare the initial state as $ \left\vert ggg,1,1\right\rangle $. 
While for \textbf{case (\emph{g})} only a single photon needs to be injected for preparing the initial state as $ \left\vert ggg,1,0\right\rangle $. And the corresponding driving Hamiltonian is almost same as Eq. \eqref{Hd}, which just replaces $ \sigma^{i}_{-} + \sigma^{i}_{+} $ with $ b+b^{\dagger} $.
In the injection process, the additional nonlinearity such as Kerr, cross-Kerr, and Pockels effects need to be taken into consideration \cite{exp102, Nori202}.

\begin{figure}[btp]
\renewcommand\thefigure{S13}
\centering
\includegraphics[width=5 in]{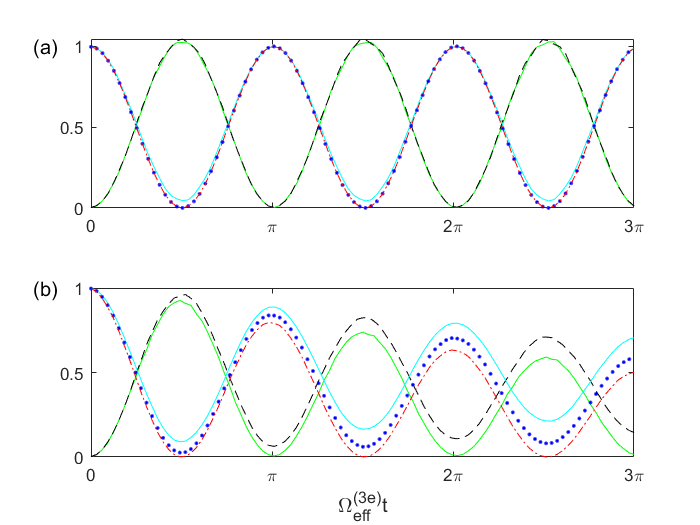}
\caption{ Time evolution of the cavity mean photon number $ \langle X^- X^+ \rangle $ (green solid curve), the ion 1 mean excitation number $ \langle C_1^- C_1^+ \rangle $ (cyan dashed curve), the zero-delay two-ion correlation function $ S^{(2)}$ (blue dotted  curve), the zero-delay three-ion correlation function $ S^{(3)}$ (red dot-dashed curve) and the mean phonon number $ \langle P^- P^+ \rangle $ (black dashed curve) by preparing the initial state of the system as $ \left\vert eee,0,0\right\rangle $. (a) System dynamics with no decay. (b) System dynamics with decay rates $ \kappa=\gamma= 2\zeta=1\times10^{-4}\omega_0 $. Other parameters are  $\eta g/\omega_{0}=0.06$ and $\nu/\omega_{0}=1.5$.}
\label{oneTE15}
\end{figure}

%\begin{figure}[bp]
%\renewcommand\thefigure{S15}
%\centering
%\includegraphics[width=4in]{TEdamp.png}
%\caption{Time evolution of the cavity mean photon number $ \langle X^- X^+ \rangle $ (blue curve), the ion 1 mean excitation number $ \langle C_1^- C_1^+ \rangle $ (black  curve), the zero-delay two-ion correlation function $ S^{(2)}$ (cyan  curve), the zero-delay three-ion correlation function $ S^{(3)}$ (red  curve) and the mean phonon number $ \langle P^- P^+ \rangle $ (magenta curve) by preparing the initial state of the system as $ \left\vert eee,0,0\right\rangle $ with decay rates $\kappa=\gamma=2\zeta=1\times 10^{-4}\omega_0  $.}
%\label{TEdamp}
%\end{figure}
%
%
%\begin{figure}[bp]
%\renewcommand\thefigure{S16}
%\centering
%\includegraphics[width=5 in]{oneTE08.png}
%\caption{Time evolution of the cavity mean photon number $ \langle X^- X^+ \rangle $ (blue solid curve), the ion 1 mean excitation number $ \langle C_1^- C_1^+ \rangle $ (black dot curve), the zero-delay three-ion correlation function $ S^{(3)}$ (red dot-dashed curve) and the mean phonon number $ \langle P^- P^+ \rangle $ (magenta dashed curve) by preparing the initial state of the system as $ \left\vert eee,0,0\right\rangle $ with decay rates $\kappa=\gamma=2\zeta=1\times 10^{-4}\omega_0  $ when $ \nu/\omega_0 =0.8$.}%, $ \omega_c$ tacking $2.1589  \omega_0$
%\label{oneTE8}
%\end{figure}

\clearpage
\subsection*{V. Preparation of N00N states and GHZ states}

By tuning the cavity frequency, we can construct different entangled states \cite{entangle02}.
The time evolutions for \textbf{case (\emph{d})}: three photons excite one ion and three phonons, is
\begin{equation}\tag{S23}
\begin{split}
\left\vert g,m,n\right\rangle \quad\longrightarrow \quad & \cos\left[\sqrt{m(m-1)(m-2)}\sqrt{(n+1)(n+2)(n+3)}V_{\mathrm{eff}}^{(3d)}t\right] \left\vert g,m,n\right\rangle \\
& -i\sin\left[\sqrt{m(m-1)(m-2)}\sqrt{(n+1)(n+2)(n+3)}V_{\mathrm{eff}}^{(3d)}t\right] \left\vert e,m-3,n+3\right\rangle, 
\end{split}
\end{equation}
\begin{equation}\tag{S24}
 \begin{split}
\left\vert e,m,n\right\rangle \quad\longrightarrow \quad & \cos\left[\sqrt{n(n-1)(n-2)}\sqrt{(m+1)(m+2)(m+3)}V_{\mathrm{eff}}^{(3d)}t\right] \left\vert e,m,n\right\rangle \\
& -i\sin\left[\sqrt{n(n-1)(n-2)}\sqrt{(m+1)(m+2)(m+3)}V_{\mathrm{eff}}^{(3d)}t\right] \left\vert g,m+3,n-3\right\rangle;
\end{split}
\end{equation} 
for \textbf{case (\emph{g})}: one photon can simultaneously excite three ions and one phonon, 
\begin{equation}\tag{S25}
\left\vert ggg,m,n\right\rangle \quad\longrightarrow \quad  \cos\left[\sqrt{m(n+1)}\Omega_{\mathrm{eff}}^{(3g)}t\right] \left\vert ggg,m,n\right\rangle  -i\sin\left[\sqrt{m(n+1)}\Omega_{\mathrm{eff}}^{(3g)}t\right] \left\vert eee,m-1,n+1\right\rangle,
\end{equation} 
\begin{equation}\tag{S26}
\left\vert eee,m,n\right\rangle \quad\longrightarrow \quad  \cos\left[\sqrt{(m+1)n}\Omega_{\mathrm{eff}}^{(3g)}t\right] \left\vert eee,m,n\right\rangle  -i\sin\left[\sqrt{(m+1)n}\Omega_{\mathrm{eff}}^{(3g)}t\right] \left\vert ggg,m+1,n-1\right\rangle.
\end{equation} 
With above time evolutions of states, we can construct the $ N=3 $  N00N state and $ \mathrm{GHZ}_{4} $ state.

\subsubsection*{A. Preparation of  $ N=3 $  $ \mathrm{N00N}$ state}

Suppose that the trapped ion is prepared in a coherent superposition of two energy eigenstates $ \left\vert \varphi\right\rangle =\cos\theta \left\vert e\right\rangle  +e^{i\phi}\sin\theta \left\vert g\right\rangle  $. The cavity state is prepared to ground state, while the motional state of the ion is prepared in 3 phonon state.
So we have the following state vector for the ion-field state
\begin{equation}\tag{S27}
 \Psi (0)=(\cos\theta \left\vert e\right\rangle  +e^{i\phi}\sin\theta \left\vert g\right\rangle  )\left\vert 0,3\right\rangle .
\end{equation} 
At a time $ t $, the ion-field state vector for \textbf{case (\emph{d})} will become
\begin{equation}\tag{S28}
\Psi (t)=\cos\theta \left[\cos(\Omega_{\mathrm{eff}}^{(3d)}t) \left\vert e,0,3\right\rangle -i\sin(\Omega_{\mathrm{eff}}^{(3d)}t)\left\vert g,3,0\right\rangle\right] +e^{i\phi}\sin\theta \left\vert g,0,3\right\rangle,
\end{equation} 
with $\Omega_{\mathrm{eff}}^{(3d)}= 6V_{\mathrm{eff}}^{(3d)} $. When interaction time satisfies $ t_k=\pi(4k+3)/2\Omega_{\mathrm{eff}}^{(3d)} $ ($k=0,1,2\cdots $), the resulting state is
\begin{equation}\tag{S29}
 \left\vert \Psi (t_k)\right\rangle   =   i\cos\theta  \left\vert g,3,0\right\rangle +e^{i\phi} \sin\theta \left\vert g,0,3\right\rangle.  
\end{equation}
If measuring the ion in its internal state $ \left\vert g\right\rangle  $ for equally weighted ionic states ($ \theta=\pi/4 $),  we get a $ N=3 $ N00N state
\begin{equation}\tag{S30}
 \left\vert \psi\right\rangle   = \dfrac{1}{\sqrt{2}} \left( i\left\vert 3,0\right\rangle +e^{i\phi} \left\vert 0,3\right\rangle \right),
\end{equation}
where the phase $ \phi $ is fully transferred from the ionic superposition states  $ \left\vert \varphi\right\rangle $ \cite{model32}.  As a source of entanglement, the N00N states are often used for performing  high-precision measurements.

\subsubsection*{B. Preparation of   $ \mathrm{GHZ}_{4} $ state}

With the time evolution of states in \textbf{case (\emph{g})}, a 4-qubit type GHZ state can be  constructed, which is highly sought for applications to quantum communication and information. 
Consider that the cavity state is prepared in a coherent superposition state $ \left\vert \varphi\right\rangle=\cos\theta \left\vert 1\right\rangle  +e^{i\phi}\sin\theta \left\vert 0\right\rangle  $. Both three trapped ions and ionic motional state are prepared to ground states. So the initial state of the system is
\begin{equation}\tag{S31}
\Psi (0)=\left\vert ggg\right\rangle (\cos\theta \left\vert 0\right\rangle  + e^{i\phi} \sin\theta \left\vert 1\right\rangle  ) \left\vert 0\right\rangle.
\end{equation} 
Then we have
\begin{equation}\tag{S32}
\Psi (t)=   e^{i\phi} \sin\theta \left\vert ggg,0,0\right\rangle+\cos\theta \left[\cos(\Omega_{\mathrm{eff}}^{(3g)}t) \left\vert ggg,1,0\right\rangle -i\sin(\Omega_{\mathrm{eff}}^{(3g)}t)\left\vert eee,0,1\right\rangle\right].
\end{equation} 
For $ t_k=\pi(4k+3)/2\Omega_{\mathrm{eff}}^{(3g)} $  ($k=0,1,2\cdots   $) and  $ \theta=\pi/4 $, 
 if one measures the cavity photon in its vacuum state, the resulting  state will become a $ \mathrm{GHZ}_{4} $ state:
\begin{equation}\tag{S33}
 \left\vert \psi\right\rangle   = \dfrac{1}{\sqrt{2}} \left( e^{i\phi} \left\vert ggg,0\right\rangle +i \left\vert eee,1\right\rangle \right).
\end{equation}

\end{widetext}


\begin{thebibliography}{99}
\bibitem {deng2} When $ \delta $ is very close to $ \nu $, but $ \delta\neq \nu $, the effective coupling rate is $\Omega^{(3a)}_{\mathrm{eff}}=-\frac{27\sqrt{6}(\omega_{0}-\delta+\nu)(\eta g)^{3}}{2(\omega_{0}-\delta)(2\omega_{0}+\delta-3\nu)(\omega_{0}-\delta+3\nu)}$.%, or  $-\frac{27\sqrt{6}(2\omega_{0}+\delta-\nu)(\eta g)^{3}}{2(\omega_{0}-\delta)(2\omega_{0}+\delta-3\nu)(2\omega_{0}+\delta+3\nu)}$ if  the initial state is $ \left\vert e,0,0\right\rangle $.
\bibitem {Heffa2} The whole completed expression of Effective Hamiltonian \eqref{Ha} should be $H_{\mathrm{eff}}=H_0+H^{(1a)}_{\mathrm{eff}}+ H^{(2a)}_{\mathrm{eff}}+H^{(3a)}_{\mathrm{eff}}$, where $ H^{(1a)}_{\mathrm{eff}}$ is the interaction Hamiltonian of form \eqref{Hi} with $ i $ only taking $1$, yielding zero after rotating wave approximation.
\bibitem {Veff3b} When $ \delta $ is very close to $ \nu $, but still not equal to $ \nu $, the effective coupling strength is $\Omega_{\mathrm{eff}}^{(3b)}=-\frac{27(\eta g)^{3}}{(\omega_{0}+3\nu-3\delta)(2\omega_{0}+3\delta-3\nu)}$.
\bibitem {Veff3c} When $ \delta $ is infinitely  approaching  to $ \nu $, but $ \delta\neq \nu $, the effective coupling strength is $\Omega_{\mathrm{eff}}^{(3c)}=-\frac{27\sqrt{6}(\omega_{0}+\delta-\nu)(\eta g)^{3}}{2(\omega_{0}+\delta)(2\omega_{0}-\delta+3\nu)(\omega_{0}+\delta-3\nu)}$. %, or $-\frac{27\sqrt{6}(2\omega_{0}-\delta+\nu)(\eta g)^{3}}{2(\omega_{0}+\delta)(2\omega_{0}-\delta+3\nu)(2\omega_{0}-\delta-3\nu)}$ if the initial state is $ \left\vert e,0,1\right\rangle $.
\bibitem {three2}  When $ \delta $ is infinitely  approaching to $ \nu $, but still not equal to $ \nu $, the effective coupling strength %for the transition between $ \left\vert ggg,1,1\right\rangle $ and $ \left\vert eee,0,0\right\rangle $ 
is $\Omega_{\mathrm{eff}}^{(3e)}=-\frac{3(\eta g)^{3}}{\omega_{0}}\left(\frac{1}{2\omega_{0}-\delta+\nu} - \frac{2}{4\omega_{0}-\delta-\nu} +\frac{2}{2\omega_{0}-\delta-\nu} -\frac{4}{4\omega_{0}-\delta+\nu} \right) $.
\bibitem {Omega3g}  When $ \delta $ is very close to  $\nu $, but not equal to $ \nu $, the effective coupling rate for the transitions between $ \left\vert ggg,1,0\right\rangle $ and $ \left\vert eee,0,1\right\rangle $ is $\Omega_{\mathrm{eff}}^{(3g)}=\frac{3(3\omega_{0}+\nu)(\delta-3\nu)(\eta g)^{3}}{\omega_{0}(\omega_{0}+\nu)(4\omega_{0}+\delta+\nu)(2\omega_{0}+\delta-\nu)}$.
\end{thebibliography}
\end{document}